\documentclass[11pt,draftclsnofoot,onecolumn]{IEEEtran}

\usepackage{amsmath,amssymb,amsfonts,mathtools,pdfsync}
\usepackage[]{graphicx}
\usepackage{sidecap}
\usepackage{amsfonts}
\usepackage{setspace, enumerate}
\usepackage{color}
\usepackage{amsmath,mathtools}
\usepackage{multirow,ifpdf,pdfpages}
\usepackage[final]{showkeys} % show ref labels in pdf
\usepackage{algpseudocode,algorithm}
\usepackage{citesort,framed,footmisc}
\usepackage[]{todonotes}
\usepackage[margin=0pt,font=small, labelfont=bf]{caption}
\usepackage{cancel,subcaption,rviewport}

\allowdisplaybreaks

\usepackage{tikz}
\usetikzlibrary{matrix,arrows}
\usetikzlibrary{decorations.pathmorphing} % noisy shapes
\usetikzlibrary{fit}					% fitting shapes to coordinates
\usetikzlibrary{backgrounds,calc}	% drawing the background after the foreground

\newcounter{mycomment}

\newif\ifannotated

\annotatedtrue

\ifannotated
\newcommand{\delete}[1]{{\color{red}{\sout{#1}}}}

\newcommand{\margincomment}[1]{\marginpar{$\Rightarrow$\color{red}\fbox{\parbox{\linewidth}{\color{black}\scriptsize#1}}}}
\else
\newcommand{\delete}[1]{{\unskip \ignorespaces}}

\newcommand{\margincomment}[1]{}
\fi

\algrenewcommand\algorithmicrequire{\textbf{Input:}}
\algrenewcommand\algorithmicensure{\textbf{Output:}}
\algrenewcommand\algorithmicrepeat{\textbf{Repeat:}}

\graphicspath{{figs/}}

\newcommand{\figwidth}{0.5\textwidth}
\newif\iffullfigure

\newcommand{\mA}{\ensuremath{\mathbf A}}

\newcommand{\mF}{\ensuremath{\mathbf F}}

\newcommand{\mW}{\ensuremath{\mathbf W}}

\newcommand{\va}{\ensuremath{\mathbf a}}

\newcommand{\vd}{\ensuremath{\mathbf d}}
\newcommand{\ve}{\ensuremath{\mathbf e}}
\newcommand{\vf}{\ensuremath{\mathbf f}}
\newcommand{\vg}{\ensuremath{\mathbf g}}

\newcommand{\vp}{\ensuremath{\mathbf p}}
\newcommand{\vq}{\ensuremath{\mathbf q}}

\newcommand{\vu}{\ensuremath{\mathbf u}}

\newcommand{\vw}{\ensuremath{\mathbf w}}
\newcommand{\vx}{\ensuremath{\mathbf x}}
\newcommand{\vy}{\ensuremath{\mathbf y}}
\newcommand{\vz}{\ensuremath{\mathbf z}}

\newcommand{\mPsi}{\ensuremath{\mathbf \Psi}}
\newcommand{\mPhi}{\ensuremath{\mathbf \Phi}}

\newcommand{\deq}{\ensuremath{ \overset{\text{def}}{=} }}

\newcommand{\<}{\langle}
\renewcommand{\>}{\rangle}

\newcommand{\delx}{\ensuremath{\partial \mathbf{x}}}

\newcommand{\amatrix}[1]{\begin{bmatrix}#1\end{bmatrix}}							 % matrix (bmatrix)
\newcommand{\sign}[1]{\mathrm{sign}{\left(#1\right)}}									 % sign function

\begin{document}
\IEEEoverridecommandlockouts
%--------------------------------------------------------------------------
% \title{Streaming Signal Recovery From Under-Sampled Measurements Using Lapped Orthogonal Transforms}
\title{Sparse Recovery of Streaming Signals Using $\ell_1$-Homotopy}
% Streaming $|ell_1$ recovery

%\author{\IEEEauthorblockN{M. Salman Asif and Justin Romberg\IEEEauthorrefmark{1}} \\
%\IEEEauthorblockA{\IEEEauthorrefmark{1}School of Electrical and Computer Engineering, Georgia Institute of Technology, Atlanta, Georgia 30332--0250\\
%Email: \{sasif, jrom\}@ece.gatech.edu} \\
%\thanks{Manuscript submitted to ...}}

\author{M.~Salman~Asif~and~Justin~Romberg
\thanks{M. S. Asif and J. Romberg are with the School of Electrical and Computer Engineering, Georgia Institute of Technology, Atlanta, GA 30332, USA. Email: \{sasif,jrom\}@gatech.edu.}
% \thanks{Manuscript submitted to the {\it IEEE Transactions on Signal Processing}, June, 2013. }
}

\maketitle

\begin{abstract}
% Motivation
% Problem statement
% Approach
% Results
% Conclusions

% Contribution: Reconstructing streaming signal from under-sampled measurements using signal representation in compact, overlapping supports.
%
% Motivation
Most of the existing methods for sparse signal recovery assume a {\it static} system: the unknown signal is a finite-length vector for which a fixed set of linear measurements and a sparse representation basis are available and an $\ell_1$-norm minimization program is solved for the reconstruction.
However, the same representation and reconstruction framework is not readily applicable in a {\it streaming} system: the unknown signal changes over time, and it is measured and reconstructed sequentially over small time intervals.
A streaming framework for the reconstruction is particularly desired when dividing a streaming signal into disjoint blocks and processing each block independently is either infeasible or inefficient.
% For instance, dividing a time-varying sinusoidal signal into disjoint blocks and using a discrete Fourier or cosine transform for sparse representation may introduce artificial discontinuities at the boundaries and ruin the sparsity.

% Problem statement
In this paper, we discuss two such streaming systems and a homotopy-based algorithm for quickly solving the associated weighted $\ell_1$-norm minimization programs:
1) Recovery of a smooth, time-varying signal for which, instead of using block transforms, we use lapped orthogonal transforms for sparse representation.
2) Recovery of a sparse, time-varying signal that follows a linear dynamic model.
%
% Approach
For both the systems, we iteratively process measurements over a sliding interval and solve a weighted $\ell_1$-norm minimization problem for estimating sparse coefficients.
Since we estimate overlapping portions of the streaming signal while adding and removing measurements, instead of solving a new $\ell_1$ program from scratch at every iteration, we use an available signal estimate as a starting point in a homotopy formulation. Starting with a warm-start vector, our homotopy algorithm updates the solution in a small number of computationally inexpensive homotopy steps as the system changes.
The homotopy algorithm presented in this paper is highly versatile as it can update the solution for the $\ell_1$ problem in a number of dynamical settings.
% We present a versatile homotopy-based algorithm that can update the solution of various $\ell_1$ problems for a number of  dynamical settings.
%
% Results
% Conclusions
%
We demonstrate with numerical experiments that our proposed streaming recovery framework outperforms the methods that represent and reconstruct a signal as independent, disjoint blocks, in terms of quality of reconstruction, and that our proposed homotopy-based updating scheme outperforms current state-of-the-art solvers in terms of the computation time and complexity.

%\begin{keywords}
%Sparse recovery, compressive sensing, lapped transform, Kalman filter, $\ell_1$ homotopy
%\end{keywords}

\end{abstract}
%
% intro,
\section{Introduction}\label{sec:intro}
% Streaming signal recovery
% Orthogonal transforms with overlapping but compact support (LOT and DWT)
%   advantages over block transform
% Linear dynamic model
% L1 homotopy (fast update)

% Motivation and overview of our work
In this paper we discuss the problem of estimating a sparse, time-varying signal from incomplete, streaming measurements---a problem that arises in a variety of signal processing applications; see  \cite{CRT06_RobustUP,Donoho_2006_CS,Candes_2006_CompressiveSampling,Baraniuk_2008_ModelCS,Olshausen-1997-sparse,Kreutz-2003-DictionaryLearning,Lustig_2008_CSMRI_SPMag,lewicki1999coding,Li2007estimation} for examples.
Most of the existing sparse recovery methods are \emph{static} in nature: they assume that the unknown signal is a finite-length vector, for which a fixed set of linear measurements and a sparse representation basis are available, and solve an $\ell_1$-norm minimization problem, which encourages the solution to be sparse while maintaining fidelity toward the measurements~\cite{Tibshirani_1996_LASSO,Chen_99_BasisPursuit,CandesTao_2007_DS}.
However, the same representation and reconstruction framework is not readily applicable in a {\it streaming} system in which the unknown signal varies over time and has no clear beginning and end. Instead of measuring the entire signal or processing the entire set of measurements at once, we perform these task sequentially over short, shifting time intervals~\cite{ARBV_ICIP2010,BA10_CISS}.

We consider the following time-varying linear observation model for a discrete-time signal $x[n]$:
\begin{equation}
y_t = \Phi_t x_t + e_t, \label{eq:LDSy}
\end{equation}
where $x_t$ is a vector that represents $x[n]$ over an interval of time, $y_t$ is a vector that contains measurements of $x_t$, $\Phi_t$ is a measurement matrix, and $e_t$ is noise in the measurements.
We use subscript $t$ to indicate that the system in \eqref{eq:LDSy} represents a small part of an infinite-dimensional streaming system, in which for any $t$, $x_t$ precedes $x_{t+1}$ in $x[n]$ and the two may overlap.
If we treat the $x_t$ independent from the rest of the streaming signal $(x[n])$, we can solve \eqref{eq:LDSy} as a stand-alone system for every $t$ as follows. Suppose we can represent each $x_t$ as $\Psi_t \alpha_t$, where $\Psi_t$ denotes a representation matrix (e.g., a discrete cosine or a wavelet transform) for which $\alpha_t$ is a sparse vector of transform coefficients. We write the equivalent system for \eqref{eq:LDSy} as
\begin{equation}\label{eq:LDSa}
y_t=\Phi_t \Psi_t \alpha_t + e_t
\end{equation}
and solve the following weighted $\ell_1$-norm minimization problem for a sparse estimate of $\alpha_t$:
\begin{equation}\label{eq:BPDN}
\underset{\alpha_t}{\text{minimize}}\; \|W_t\alpha_t\|_1 + \frac{1}{2}\|\Phi_t \Psi_t \alpha_t -y_t\|_2^2.
\end{equation}
The $\ell_1$ term promotes sparsity in the estimated coefficients; $W_t$ is a diagonal matrix of positive weights that can be adapted to promote a certain sparse structure in the solution~\cite{zou2006adaptive,Candes_rwtL1_2008}; and the $\ell_2$ term ensures that the solution remains close to the measurements.
The optimization problem in \eqref{eq:BPDN} is convex and can be solved using a variety of solvers~\cite{Boyd_book_ConvexOptimization,Wright-2009-sparsa,yang-2011-yall1,Becker_2011_NESTA,beck-2009-FISTA}.
% Prior knowledge about the sparsity of $\vx$ helps in its recovery from the compressed measurements in \eqref{eq:y=Phix}.

\begin{figure*}
%\centering
    \begin{subfigure}[t]{0.5\textwidth}
            % \captionsetup{name=Figure}
        \centering
        \includegraphics[page=19,trim = 0mm 20mm 25mm 0mm, clip, width=1\textwidth, keepaspectratio]{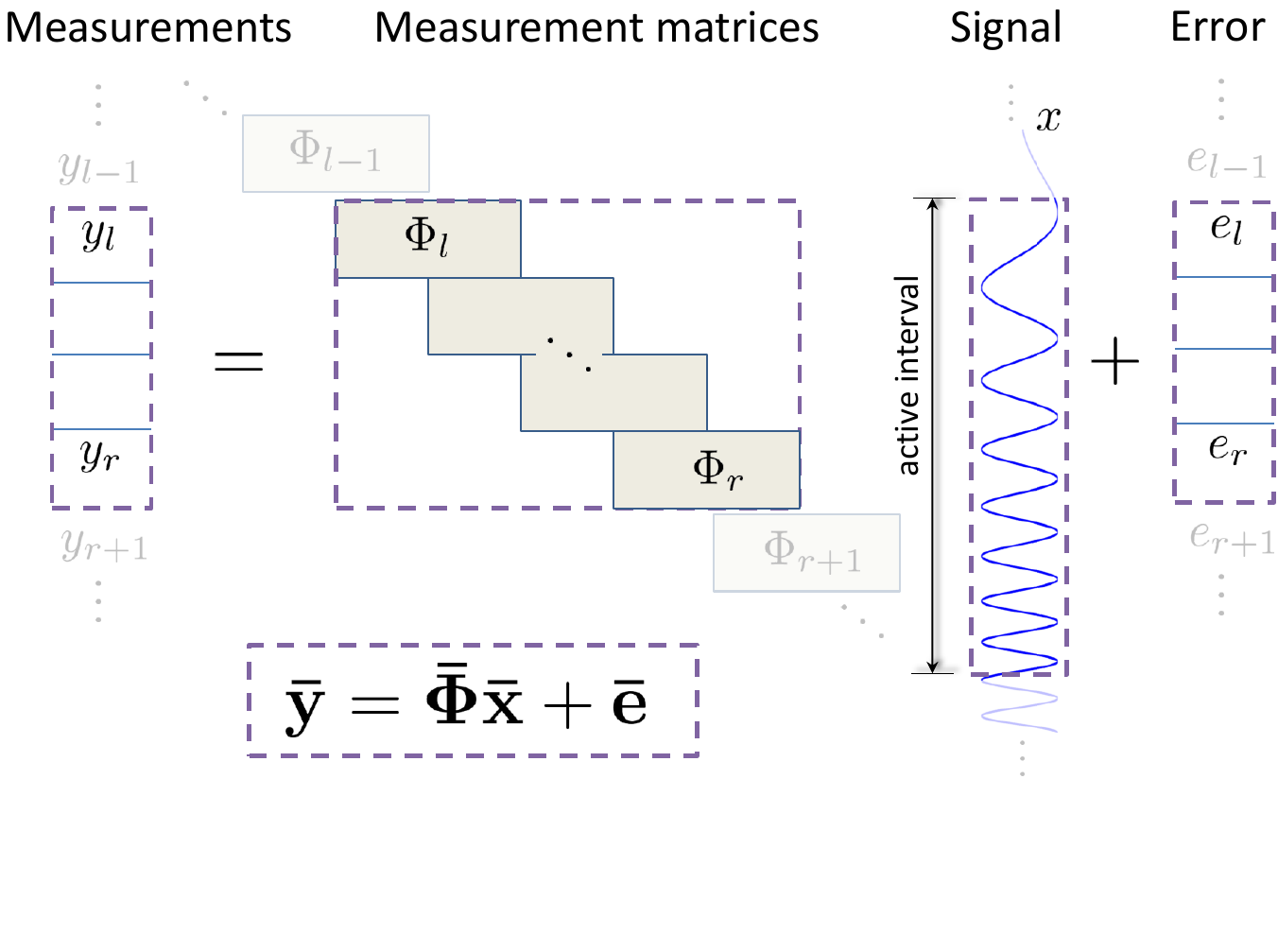}
        \caption{Overlapping measurements of a signal. }
        \label{fig:y=Phix}
    \end{subfigure}
    \hspace{5pt}
    \vline \vline
    \hspace{5pt}
    \begin{subfigure}[t]{0.50\textwidth}
        % \captionsetup{name=Figure}
        \centering
        \includegraphics[page=18,trim = 0mm 30mm 40mm 0mm, clip, width=1\textwidth, keepaspectratio]{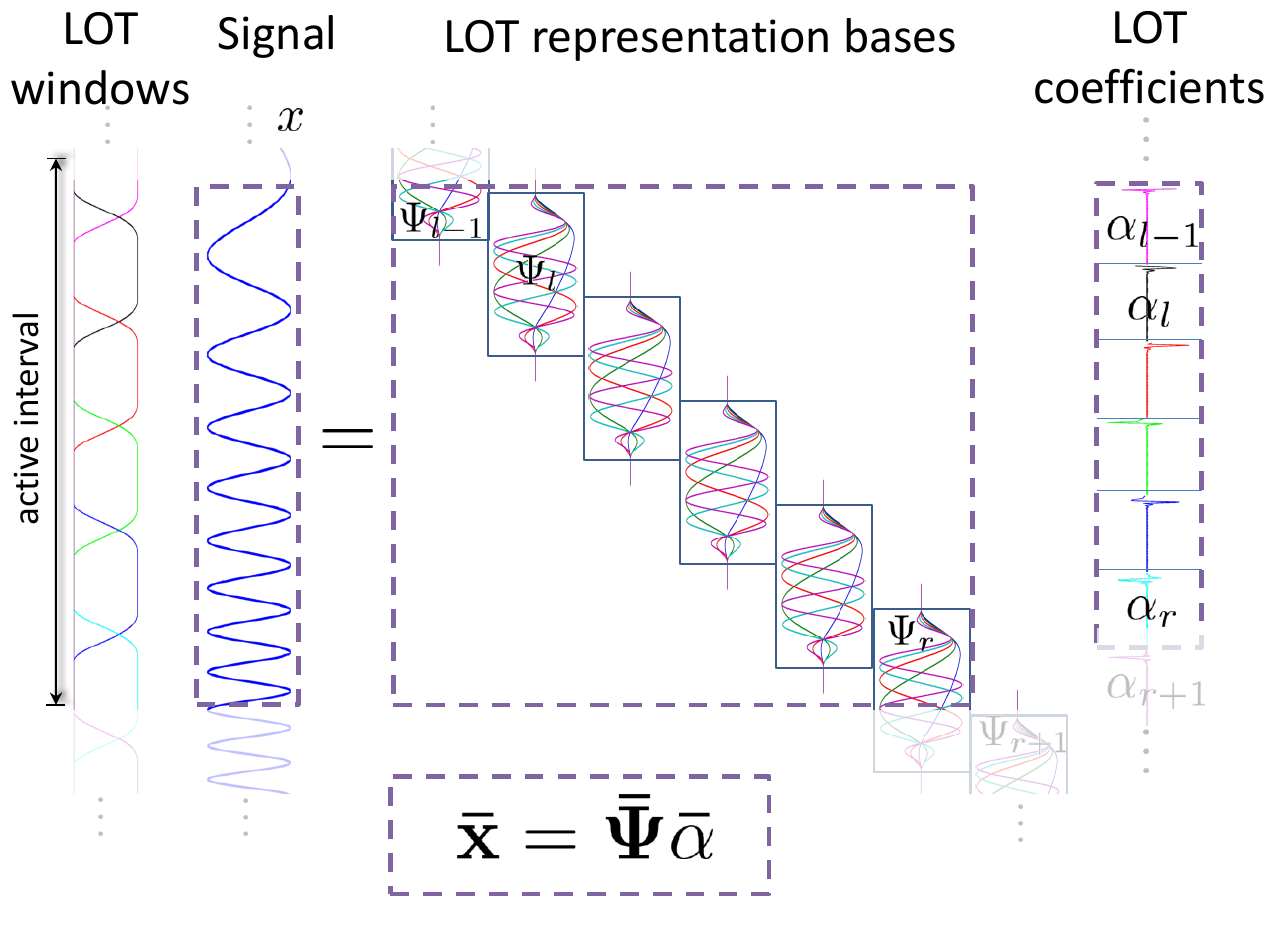}
        \caption{LOT-based representation of a signal.}
        \label{fig:x=Psia}
    \end{subfigure}
    \caption[Illustration of overlapping systems over an active interval]{Illustration of an overlapping measurement system \textbf{(a)} and a lapped orthogonal transform (LOT)-based representation system \textbf{(b)}. Boxed regions represent the system over the active interval. Subscripts $l$ and $r$ indicate the left and the right border of the active interval.}
    \label{fig:overlappingsystem}
\end{figure*}

The method described above represents and reconstructs the signal blocks $(x_t)$ independently, which is natural if both the measurement system in \eqref{eq:LDSy} and the representation system in \eqref{eq:LDSa} are block-diagonal; that is, the $x_t$ are non-overlapping in \eqref{eq:LDSy} and each $x_t$ is represented as a sparse vector using a block transform in \eqref{eq:LDSa}. However, estimating the $x_t$ independently is not optimal if the streaming system for \eqref{eq:LDSy} or \eqref{eq:LDSa} is not block diagonal, which can happen if $\Phi_t$, $\Psi_t$, or both of them overlap across the $x_t$. An illustration of such an overlapping measurement and representation system is presented in Fig.~\ref{fig:overlappingsystem}. Figure~\ref{fig:y=Phix} depicts a measurement system in which the $\Phi_t$ overlap (the $x_t$, which are not labeled in the figure, are the overlapping portions of $x[n]$ that constitute the $y_t$). Figure~\ref{fig:x=Psia} depicts a representation of $x[n]$ using lapped orthogonal transform (LOT) bases~\cite{Malvar-1989-LOT,Mallat_book_WaveletTour99} in which the $\Psi_t$ overlap (and multiple $\Psi_t\alpha_t$ may add up to constitute a portion of $x[n]$).

In this paper we present $\ell_1$-norm minimization based sparse recovery algorithms for the following two types of overlapping, streaming systems:
\begin{enumerate}[\bf 1. ]
\item
Recovery of smooth, time-varying signals from streaming measurements in \eqref{eq:LDSy} using sparse representation bases with compact but overlapping supports.
\item
Recovery of time-varying signals from streaming measurements in \eqref{eq:LDSy} in the presence of the following linear dynamic model:
\begin{equation}
x_{t+1} = F_t x_t + f_t, \label{eq:LDSx}
\end{equation}
where $F_t$ is a prediction matrix that couples $x_t$ and $x_{t+1}$ and $f_t$ is the error in the prediction, which we assume has a bounded $\ell_2$ norm.
\end{enumerate}
In both these systems, we assume that sets of measurements are sequentially recorded over short, shifting (possibly overlapping) intervals of the streaming signal according to the system in \eqref{eq:LDSy}.
Instead of estimating each block ($x_t$) independently, we iteratively estimate the signal ($x[n]$) over small, sliding intervals, which allows us to link together the blocks that share information.
At every iteration, we build a system model that describes the measurements and the sparse coefficients of the streaming signal over an active interval (one such example is depicted in Fig.~\ref{fig:overlappingsystem}).
We estimate the sparse coefficients for the signal over the active interval by solving a weighted $\ell_1$-norm minimization problem, and then we shift the active interval by removing the oldest set of measurements from the system and adding a new one.
For instance, to jointly solve the systems in \eqref{eq:LDSy} and \eqref{eq:LDSx} over an active interval that includes $P$ instances of $x_t$, say $x_1,\ldots,x_P$, which are non-overlapping and represented as $x_t=\Psi_t \alpha_t$, we solve the following modified form of \eqref{eq:BPDN}:
\begin{equation}\label{eq:BPDN2}
\underset{\alpha_1,\ldots,\alpha_P}{\text{minimize}}\; \sum_{p=1}^P \|W_p\alpha_p\|_1 + \frac{1}{2}\|\Phi_p \Psi_p \alpha_p -y_p\|_2^2 + \frac{\lambda_p}{2}\|F_{p-1} \Psi_{p-1} \alpha_{p-1} - \Psi_p \alpha_p\|_2^2,
\end{equation}
where the $\lambda_p>0$ denote the regularization parameters. To separate $\alpha_0$ from the active system, we fix its value to an estimate $\widehat \alpha_0$. At the next streaming iteration, we may have $x_2,\ldots,x_{P+1}$ in the active interval and an estimate of all but $x_{P+1}$.
However, before solving the optimization problem, an estimate of the signal over the entire active interval can be predicted. We use the available signal estimate to aide the recovery process in two ways: We update the $W_t$ using available estimates of the sparse coefficients $\alpha_t$ (in the same spirit as iterative reweighting~\cite{Candes_rwtL1_2008}), and we use the available estimates of the $\alpha_t$ as a starting point to expedite the solution of the $\ell_1$ problem.

One key contribution in this paper is a homotopy algorithm for quickly solving the weighted $\ell_1$-norm minimization problems for the streaming signal recovery; see \cite{OsbornePresnell_2000_NewApproachLasso,Efron_2004_LARS} for the background on the LASSO homotopy.
Since we sequentially estimate overlapping components of a streaming signal while adding and removing measurements, instead of solving a new optimization problem at every iteration, we use the existing signal estimate as a starting point (warm-start) in our homotopy formulation.
The homotopy algorithm that we present in this paper extends and unifies previous work on dynamic $\ell_1$ updating~\cite{Garrigues_2008_NIPS,AR_L1filtering_Asilomar08,AR_DynamicX_CISS09,AR_L1updating_JSTSP09,AR_asilomar_2010,Yang_2012_warmL1,AR_2012_rwtL1}:
Our newly proposed $\ell_1$ homotopy algorithm can quickly update the solution of the $\ell_1$ programs of the form \eqref{eq:BPDN} for arbitrary changes in $y_t,W_t, \Phi_t, \Psi_t$. For instance, adding or removing sequential measurements~\cite{Garrigues_2008_NIPS,AR_L1filtering_Asilomar08,AR_L1updating_JSTSP09,AR_asilomar_2010}, changes in the measurements ($y_t$) as the signal ($x_t$) changes with a fixed measurement matrix~\cite{AR_DynamicX_CISS09,AR_L1updating_JSTSP09}, arbitrary changes in the measurement matrix ($\Phi_t$) or the representation matrix ($\Psi_t$)~\cite{Yang_2012_warmL1,AR_asilomar_2010}, or changes in the weights ($W_t$)~\cite{AR_2012_rwtL1}.
Unlike previous approaches, we do not impose any restriction on the warm-start vector to be a solution of an $\ell_1$ problem; as we can accommodate an arbitrary vector to initialize the homotopy update. Of course, the update will be quicker when the starting vector is close to the final solution.
For solving a weighted $\ell_1$ program at every streaming iteration, we will use a thresholded and transformed version of the solution from the previous iteration as a warm-start in a homotopy program. We will sequentially add and remove multiple measurements in the system, update the weights in the $\ell_1$ term, and update the solution in a sequence of a small number of computationally inexpensive homotopy steps.

The problem formulations and the recovery methods presented in this paper compare with some of the existing signal estimation schemes.
Recursive least squares (RLS) and the Kalman filter are two classical estimation methods that are oblivious to any sparse structure in the signal and solve the systems in \eqref{eq:LDSy} and \eqref{eq:LDSx} in the least-squares sense~\cite{Golub_1996_MatrixComputation,Kalman-1960,Sorenson_GausstoKalman_1970}. One attractive feature of these methods is that their solutions admit closed form representation that can be updated recursively.
The homotopy algorithm solves an $\ell_1$ problem and its updating scheme is not as simple as that of the RLS and the Kalman filter, but the update has the same recursive spirit and reduces to a series of low-rank updates.
%
% Moreover, under the so-called Gauss-Markov model assumptions,
Moreover, under certain conditions, the recursive estimate of the standard Kalman filter that is computed using only a new set of measurements, a previous estimate of the signal, and the so-called information matrix is optimal for all the previous measurements---as if it were computed by solving a least-squares problem for all the measurements simultaneously~\cite{Kalman-1960,Kailath-2000-linear}.
In contrast, the signal estimate for the $\ell_1$ problems (such as \eqref{eq:BPDN2}) is optimal only for the measurements available in the system over the active signal interval.

Sparse recovery of smooth, time-varying signal from streaming, overlapping measurements has been discussed in \cite{BA10_CISS,BA10_MILCOM}, but the sparse recovery algorithm used there is a streaming variant of a greedy matching pursuit algorithm~\cite{Needell_2008_CoSaMP}.
Recently, several methods have been proposed to incorporate signal dynamics into the sparse signal estimation framework \cite{vaswani2008kalman,Carmi:2010,angelosante2009LassoKalman,CARR-2011-CISS,ACRR_KalmanL1L1_2011,ziniel2010tracking,AHBR_2012_MASTeR}.
The method in \cite{vaswani2008kalman} identifies the support of the signal by solving an $\ell_1$ problem and modifies the Kalman filter to estimate the signal on the identified support; \cite{Carmi:2010} embeds additional steps within the original Kalman filter algorithm for promoting a sparse solution; \cite{ziniel2010tracking} uses a belief propagation algorithm to identify the support and update the signal estimate; \cite{CARR-2011-CISS} compares different types of sparse dynamics by solving an $\ell_1$ problem for one signal block; \cite{ACRR_KalmanL1L1_2011} assumes that the prediction error is sparse and jointly estimates multiple signal blocks using a homotopy algorithm; and  \cite{angelosante2009LassoKalman} solves a group-sparse $\ell_1$ problem for a highly restrictive dynamic signal model in which locations of nonzero components of the signal do not change.
In our problem formulation, we consider a general dynamic model in \eqref{eq:LDSx} and solve a problem of the form in \eqref{eq:BPDN2} over a sliding, active interval. Our emphasis is on an efficient updating scheme for moving from one solution to the next as new measurements are added and old ones are removed.

The paper is organized as follows. We discuss signal representation using bases with overlapping supports in Section~\ref{sec:signal}, the recovery framework for the two systems in Section~\ref{sec:recovery}, and the homotopy algorithm in Section~\ref{sec:l1homotopy}. We present experimental results to demonstrate the performance of our algorithms, in terms of the quality of reconstructed signals and the computational cost of the recovery process, in Section~\ref{sec:exp}.

% \input{relatedBG}
% representation setup
\section{Signal representation using compactly supported bases}\label{sec:signal}
We will represent a discrete-time signal $x[n]$ as
\begin{equation}\label{eq:x[n]_general}
x[n] = \sum_{p\in\mathbb{Z}} \sum_{0\le k < l_p} \alpha_{p,k} \psi_{p,k}[n],
\end{equation}
where the set of functions $\psi_{p,k}$ forms an orthogonal basis of $\ell_2(\mathbb{Z})$ and the $\alpha_{p,k}=\<\psi_{p,k},x\>$ denote the corresponding basis coefficients that we expect to be sparse or compressible. For a fixed $p\in\mathbb{Z}$, $\{\psi_{p,k}\}_k$ denotes a set of orthogonal basis vectors that have a compact support over an interval $I_p$.
The supports of the $\psi_{p,k}$ and the $\psi_{p',k}$ (i.e., $I_p$ and $I_{p'}$) may overlap if $p\ne p'$. An example of such a signal representation using lapped orthogonal bases is depicted in Fig.~\ref{fig:x=Psia}, where a $\Psi_p$ denotes the basis functions in $\{\psi_{p,k}\}_k$ supported on $I_p$, an $\alpha_p$ denotes the respective $\{\alpha_{p,k}\}_k$, and the overlapping windows denote the intervals $I_p$.

While we assume a general framework for the signal representation using \eqref{eq:x[n]_general} in the derivations of the recovery problems, we use the lapped orthogonal transform (LOT)~\cite{Malvar-1989-LOT} in most of our explanations and numerical experiments.
A LOT decomposes a signal into orthogonal components with compact, overlapping supports. Orthogonality between the components in the overlapping regions is maintained due to projections with opposite (i.e., even and odd) symmetries. LOT basis functions can be designed using modified cosine-IV basis functions that are multiplied by smooth, overlapping windows.
The advantage of using the LOT instead of a simple block-based discrete cosine or Fourier transform is that block-based transforms use rectangular windows to divide a signal into disjoint blocks and that can introduce artificial discontinuities at the boundaries of the blocks and ruin the sparsity~\cite{Mallat_book_WaveletTour99}.

A discrete LOT basis can be designed as follows. Divide the support of the signal into consecutive, overlapping intervals $I_p = [a_p - \eta_p, a_{p+1}+\eta_{p+1}]$, where $\{a_p\}_{p\in \mathbb{Z}}$ is a sequence of half integers (i.e., $a_p + 1/2 \in \mathbb{Z}$) and $\{\eta_p\}_{p\in \mathbb{Z}}$ is a sequence of transition width parameters such that \mbox{$l_p \deq  a_{p+1}-a_p \ge \eta_p+\eta_{p+1}$}.
The LOT basis function, $\psi_{p,k}$ in \eqref{eq:x[n]_general}, for every $p,k$ is defined as
\begin{equation}\label{eq:psi_LOT}
\psi_{p,k}[n] = g_p[n] \sqrt{\frac{2}{l_p}} \cos\left[\pi\left(k+\frac{1}{2}\right) \frac{n-a_p}{l_p} \right],
\end{equation}
which is a translated and dilated cosine-IV basis function, multiplied by a smooth window $g_p$ that is supported on $I_p$.
For a careful choice of $g_p$, coupled with the even and odd symmetry of cosine-IV basis functions with respect to $a_p$ and $a_{p+1}$, respectively, the set of functions $\psi_{p,k}$ forms an orthonormal basis of $\ell_2(\mathbb{Z})$ (see \cite[Sec.\ 8.4]{Mallat_book_WaveletTour99} for further details).

To compute the LOT coefficients of $x[n]$ over an arbitrary interval $\Pi$, we assume a partition of $\Pi$ into appropriate LOT subintervals $I_p$.
Figure~\ref{fig:LOT_projections} depicts an example with such a partition of a time interval using LOT windows and the decomposition of a linear chirp signal into overlapping components using LOT bases and their respective coefficients.
Since the set of functions $\{\psi_{p,k}\}_k$ defines an orthogonal basis for a LOT subspace on respective $I_p$, the corresponding LOT projection of $x[n]$ can be written as
\begin{equation}\label{eq:xt_p1}
\widetilde{x}_p[n] = \sum_{k=0}^{l_p-1}\underbrace{\<x,\psi_{p,k}\>}_{\alpha_{p,k}}\psi_{p,k}[n],
\end{equation}
where $\widetilde x_p[n]$ is supported on $I_p$.
We can represent the restriction of $\widetilde{x}_p[n]$ on $I_p$ as $\Psi_p \alpha_p$,
where $\Psi_p$ is a synthesis matrix whose $k^\textrm{th}$ column consists of $\psi_{p,k}[n]$ restricted to $I_p$ and $\alpha_p$ is an $l_p$-length vector of LOT coefficients that consists of $\{\alpha_{p,k}\}_k$ for ${0\le k< l_p}$.
Note that the $\widetilde x_p[n]$ are the overlapping, orthogonal components of $x[n]$, and to synthesize $x[n]$ over $\Pi$, we have to add all the $\widetilde x_p[n]$ that overlap $\Pi$.
Referring to Fig.~\ref{fig:x=Psia}, $\bar \vx$ denotes $x[n]$ over the active interval $\Pi$, $\bar \alpha$ denotes a vector that contains all the $\alpha_p$ that contribute to $\bar \vx$ stacked on top of one another,
$\bar \mPsi$ contains the corresponding $\Psi_p$ (in part or full) at appropriate columns and rows, and the columns of
$\Psi_p$ and $\Psi_{p+1}$ overlap in $2\eta_{p+1}$ rows.

Another example of an orthogonal basis that can be naturally separated into overlapping, compact intervals is the wavelet transform. A wavelet transform decomposes a signal into orthogonal components with overlapping supports at different resolutions in time and frequency~\cite{Vetterli_Book_WaveletsAndSubband,Mallat_book_WaveletTour99}. The scaling and wavelet functions used for this purpose overlap one another while maintaining orthogonality. Although commonly used filter-bank implementations assume that the finite-length signals are symmetrically or periodically extended during convolution, which yields a block-based wavelet transform, we can write wavelet bases in terms of shifted, dilated wavelet and scaling functions that overlap across adjacent blocks.
Figure~\ref{fig:DWT_projections} depicts an example of the decomposition of a piece-wise smooth signal into overlapping components using wavelet bases and their respective coefficients.

% \fullfiguretrue
% \fullfigurefalse
\renewcommand{\figwidth}{0.475\textwidth}
\begin{figure*}
    \begin{subfigure}[t]{\figwidth}
        % \captionsetup{name=Figure}
        \centering
        \includegraphics[page=1,trim = 0mm 0mm 0mm 7mm, clip,  width=1\columnwidth, height = 5cm]{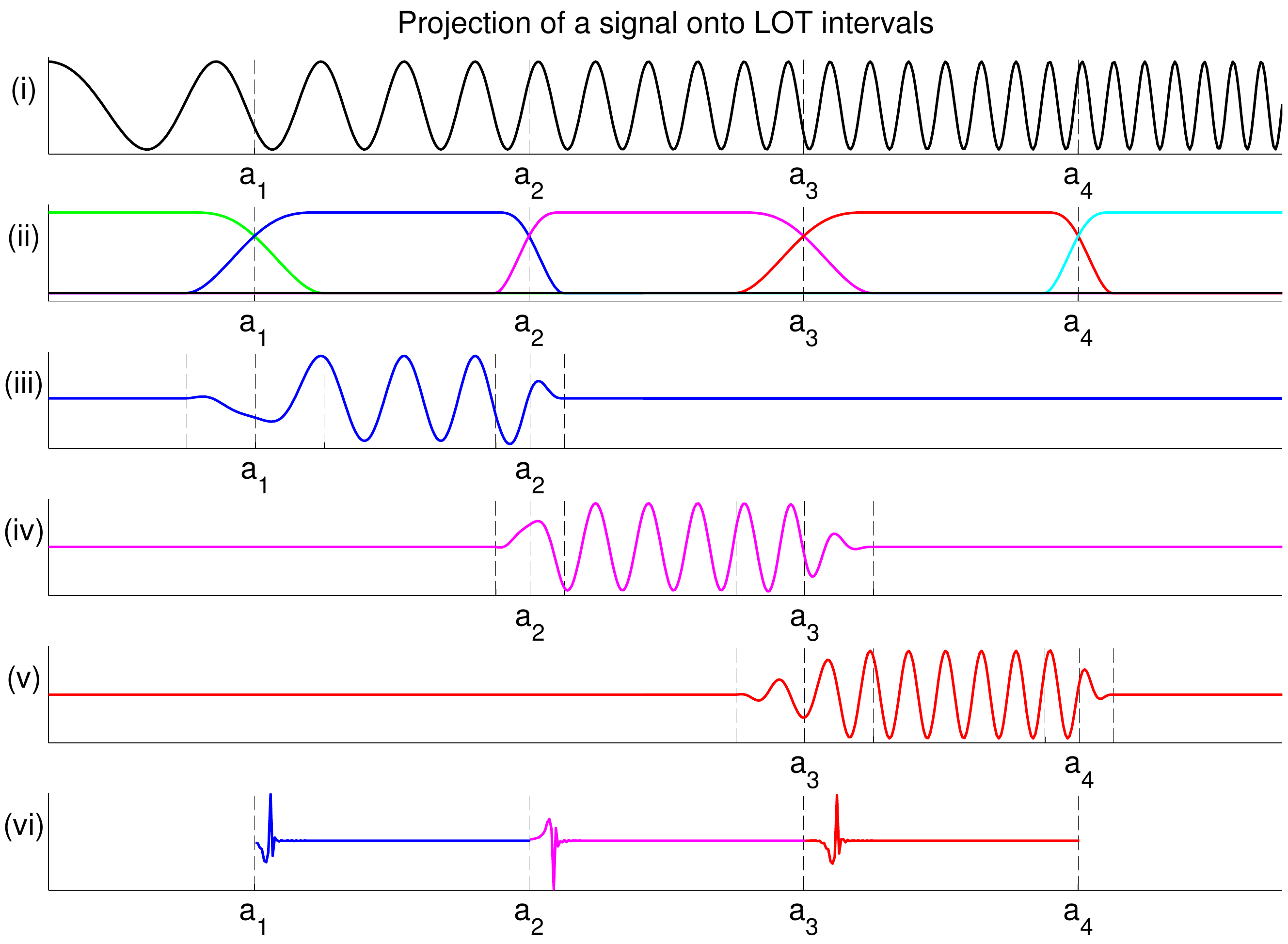}
        \caption{LOT projections and coefficients. (i) A discrete-time linear chirp signal ($x[n]$). (ii) LOT windows over different intervals $(I_p)$; distance between dotted lines around $a_p$ represent $\eta_p$. (iii--v) LOT projections $\tilde x_p[n]$ over respective intervals. (vi) Sparse coefficients ($\alpha_{p,k}$).}
        \label{fig:LOT_projections}
    \end{subfigure}
    \hspace{10pt}
            \begin{subfigure}[t]{\figwidth}
            % \captionsetup{name=Figure}
        \centering
        \includegraphics[page=1,trim = 0mm 0mm 0mm 7mm, clip,  width=1\columnwidth, height = 5cm]{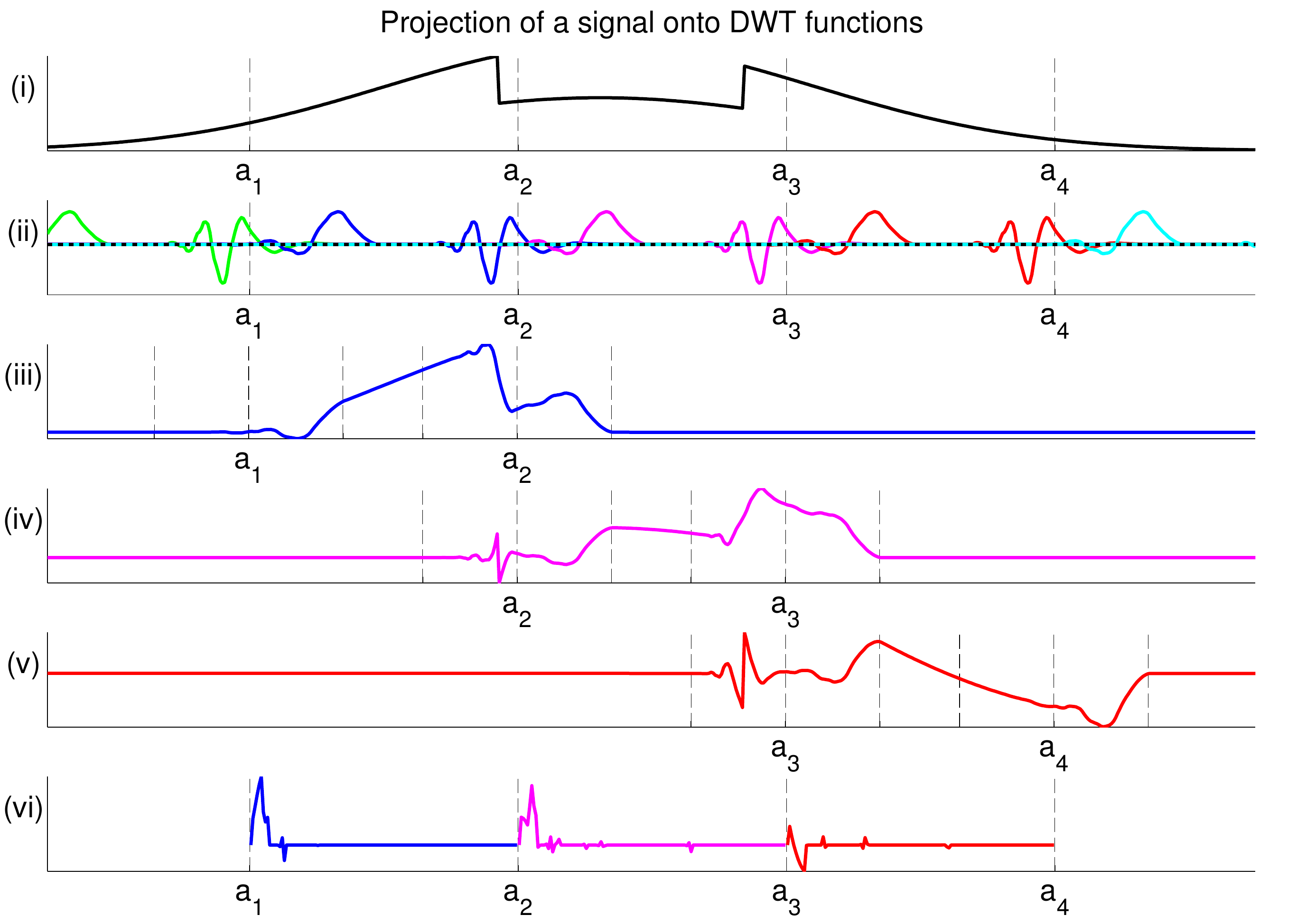}
        \caption{Wavelet projections and coefficients. (i) A piecewise smooth signal ($x[n]$). (ii) A subset of scaling and wavelet functions at the coarsest scale. Dotted lines denote $I_p$. (iii--v) Wavelet projections $\tilde x_p[n]$ over respective intervals. (vi) Sparse coefficients ($\alpha_{p,k}$).}
        \label{fig:DWT_projections}
    \end{subfigure}
        \caption[Signal decomposition in LOT and wavelet bases.]{Signal decomposition in \textbf{(a)} LOT and \textbf{(b)} wavelet bases.}
    \label{fig:projections}
\end{figure*}

% measurement setup
% \input{sensing-model}
% recovery setup
\section{Sparse signal recovery from streaming measurements}\label{sec:recovery}
In a streaming system, we iteratively estimate sparse coefficients of the signal over an active, sliding interval. We describe a system for the measurements and the sparse representation of the signal over the active interval and solve a weighted $\ell_1$-norm minimization problem for estimating the sparse coefficients.
At every iteration of the streaming recovery process, we shift the active interval by removing a few oldest measurements and adding a few new ones in the system. Estimate of the sparse coefficients and the signal portion that leave the active interval are committed to the output. The length of the active interval determines the delay, memory, and computational complexity of the system.

% \subsection{System over active interval}
Consider the linear system in \eqref{eq:LDSy}: $y_t = \Phi_t x_t + e_t$, where $x_t$ denotes a portion of $x[n]$ over a short interval and the consecutive $x_t$ may also overlap.
% We assume that the active interval $\Pi$ consists of a small number of sampling intervals $\Omega_t$.
We denote $x[n]$ over the active interval $\Pi$ as $\bar \vx$ and assume that $\bar \vx$ consists of a small number of $x_t$. We describe the equivalent system for $\bar \vx$ in the following compact form:
\begin{equation}\label{eq:y=Phix_bar}
\bar \vy = \bar \mPhi \bar \vx + \bar \ve,
\end{equation}
where $\bar \vy$ denotes a vector that contains $y_t$ for the $x_t$ that belong to $\bar \vx$, $\bar \mPhi$ denotes a matrix that contains the corresponding $\Phi_t$, and $\bar \ve$ denotes the noise vector. At every iteration of the streaming recovery algorithm, we shift $\Pi$ by removing the oldest $y_t$ in the system and adding a new one and update the system in \eqref{eq:y=Phix_bar} accordingly. An example of such a measurement system in depicted in Fig.~\ref{fig:y=Phix}, where the active system is represented in a boxed region.
To represent the signal $\bar \vx$ using the model in \eqref{eq:x[n]_general}, we use the following compact form:
\begin{equation}\label{eq:x=Psia_bar}
\bar \vx = \bar \mPsi \bar \alpha,
\end{equation}
where $\bar \alpha$ contains the $\alpha_{p,k}$ that synthesize $\bar \vx$ and the synthesis matrix $\bar \mPsi$ contains the corresponding $\psi_{p,k}$ restricted to $\Pi$ as its columns. An example of such a representation system in depicted in Fig.~\ref{fig:x=Psia}.

In the following two sections, we discuss the problem formulation for the recovery of a streaming signal from streaming measurements when 1) the signal is represented using lapped orthogonal bases and 2) the signal changes according to a linear dynamic model.

%
%1. System model setup, representation basis aligned with measurements so that better/sparse representation is achieved
%2. We use solution of previous iteration to predict a warm-start for the current iteration,
%3. Weighted l1 norm solution using warm-start methods: Since a significant portion of the signal overlaps from one iteration to the next, we use a warm-start method to quickly solve the problem.
%

%----------%
% LOT
%----------%
%
\subsection{Streaming signal with lapped orthogonal bases}\label{sec:streaming_LOT}
% Measurements: \bar \vy = \bar \mPhi \bar \vx + \bar \ve
% Representation: \bar \vx = \bar \mPsi \bar \alpha
\subsubsection{System model}\label{sec:model_LOT}
\begin{figure*}
% \centering
    \begin{subfigure}[t]{0.47\textwidth}
        % \captionsetup{name=Figure}
        \centering
        \includegraphics[page=20,trim = 0mm 60mm 20mm 0mm, clip, width=1\textwidth, keepaspectratio]{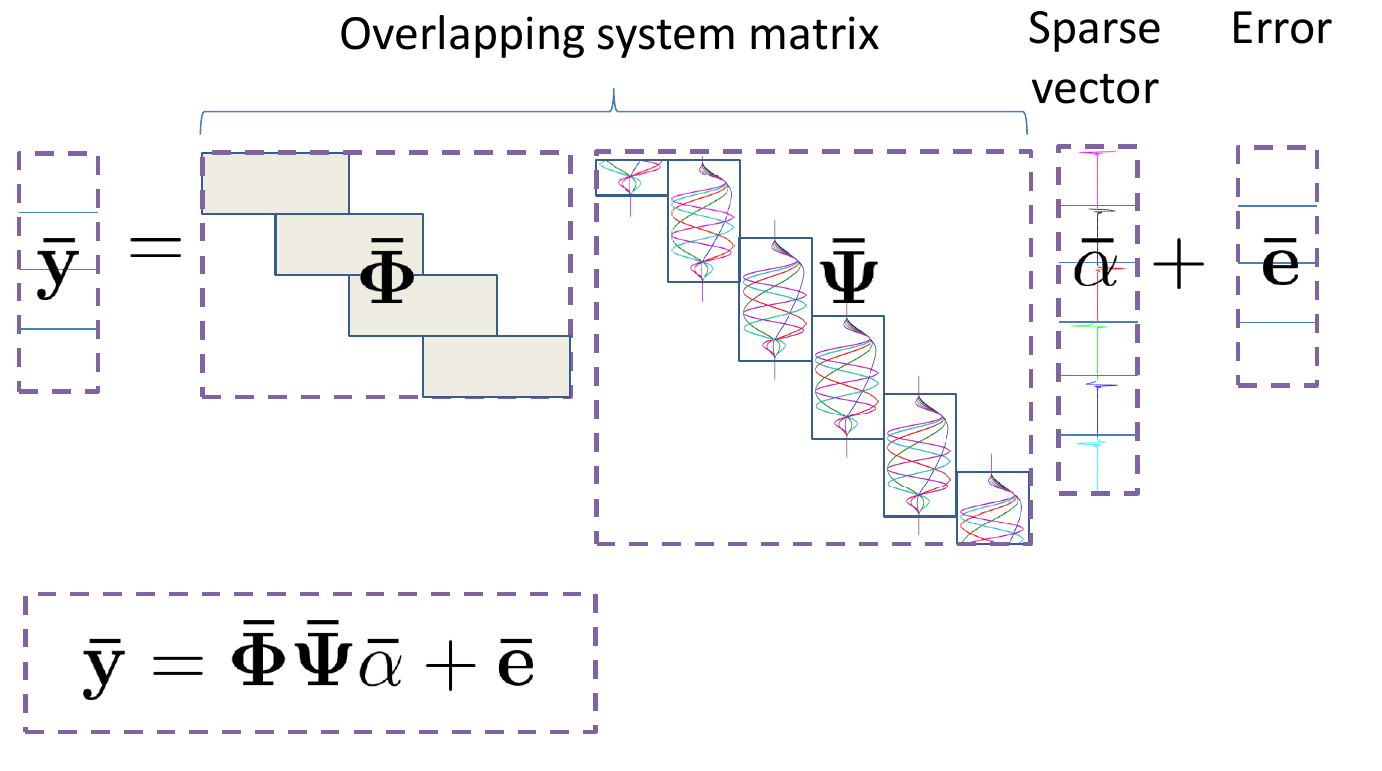}
        \caption{}%{System over the active interval.}
        \label{fig:y=PhiPsia}
    \end{subfigure}
    \hspace{5pt}
    \vline\vline
    \hspace{5pt}
    % $\Rightarrow$
    \begin{subfigure}[t]{0.51\textwidth}
            % \captionsetup{name=Figure}
        \centering
        \includegraphics[page=21,trim = 0mm 60mm 0mm 0mm, clip, width=1\textwidth, keepaspectratio]{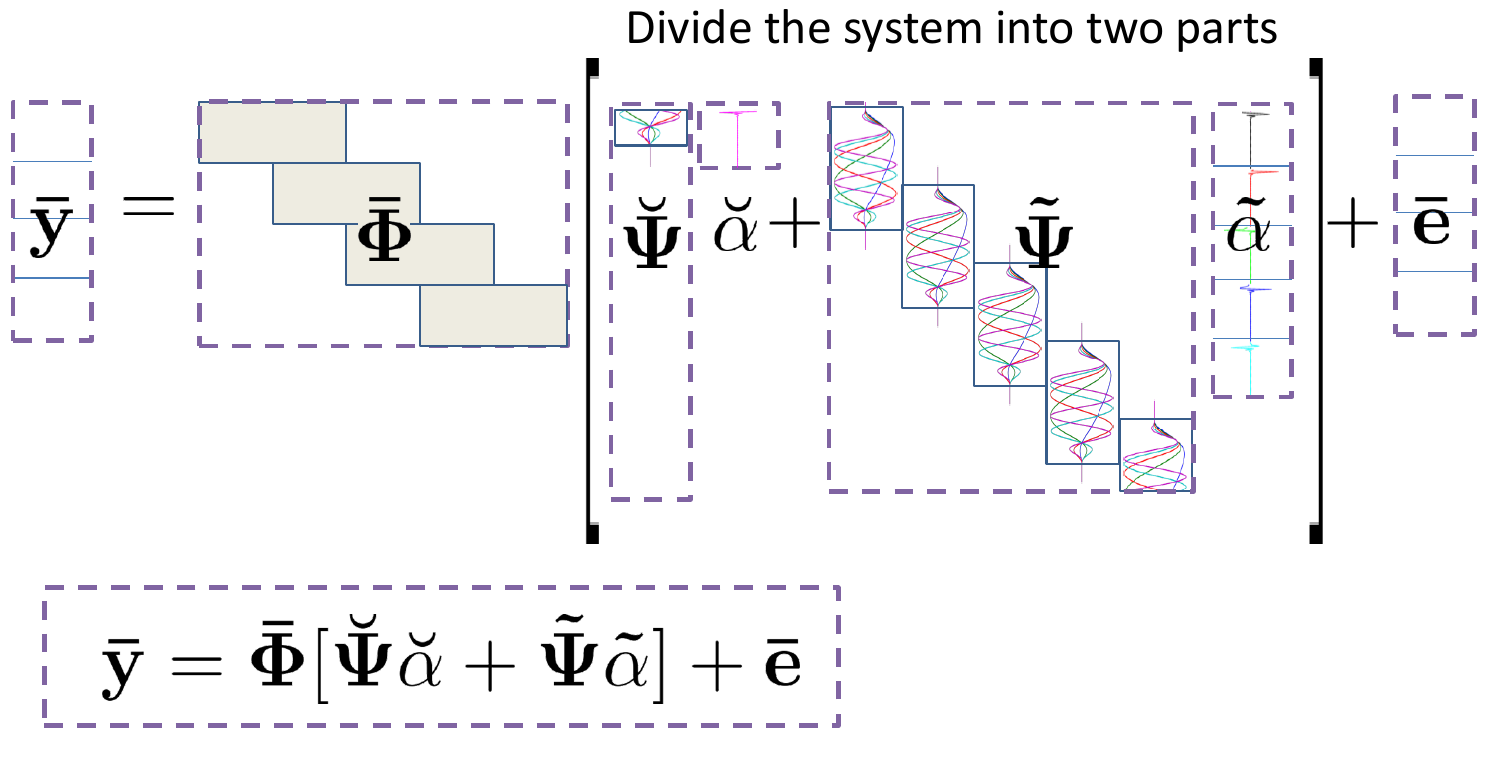}
        \caption{}% {System divided into two parts.}
        \label{fig:yt=PhiPsiat}
    \end{subfigure}
    \caption[Illustration of the system used for the signal reconstruction.]{Illustration of the system used for the signal reconstruction. \textbf{(a)} System over the active interval. \textbf{(b)} System divided into two parts so that $\breve \alpha$ can be removed.}
    \label{fig:activesystem}
\end{figure*}

Given the system in \eqref{eq:y=Phix_bar} for active interval $\Pi$, we use lapped orthogonal bases for signal representation in \eqref{eq:x=Psia_bar} and describe the system in the following equivalent form:
\begin{equation}\label{eq:y=PhiPsia_bar}
\bar \vy = \bar \mPhi \bar \mPsi \bar \alpha + \bar \ve.
\end{equation}
An example of such a system is depicted in Fig.~\ref{fig:y=PhiPsia}.
Note that even when $\bar \mPhi$ is a block diagonal matrix, the system in \eqref{eq:y=PhiPsia_bar} cannot be separated into independent blocks if $\bar \mPsi$ has overlapping columns.
%
% \Salnote{Even if $\bar \mPhi$ is block-diagonal, the combined system matrix $\bar \mPhi \bar \mPsi$ is not separable...}

One important consideration in our system is the design of $\bar \mPsi$ with respect to the decomposition of $\Pi$ into overlapping intervals $I_p$. Our motivation is to have as few unknown coefficients in $\bar \alpha$ as possible. Note that if an interval $I_p$ overlaps with $\Pi$ (partially or fully), we have to include its corresponding coefficient vector $\alpha_p$ of length $l_p$ into $\bar \alpha$.
Since we can divide the interior of $\Pi$ in an arbitrary fashion, the special consideration is only for the $I_p$ that partially overlap with $\Pi$ on its left and right borders.

On the right end of $\Pi$, we align the right-most interval, say $I_r$, such that it partially overlaps with $\Pi$ but the interval after that, say $I_{r+1}$,  lies completely outside $\Pi$. In such a case $\bar \alpha$ would contain $\alpha_r$ but not $\alpha_{r+1}$. Such a relationship between the active interval $(\Pi)$ and the subintervals $(I_p)$ is depicted in Fig.~\ref{fig:x=Psia}, where we adjusted the right-most interval such that the overlapping region on its right side lies outside the active interval $\Pi$.

On the left end of $\Pi$, we align the left-most interval, say $I_l$, such that it is fully included in $\Pi$. However, in such a setting $I_{l-1}$ will partially overlap with $\Pi$ and the corresponding coefficient vector $\alpha_{l-1}$ of length $l_{l-1}$ will be included in $\bar \alpha$. Suppose we have committed the estimate of $\alpha_{l-1}$ to the output, and we want to remove it from the system in \eqref{eq:y=PhiPsia_bar}. If the system in \eqref{eq:y=PhiPsia_bar} were block-diagonal, we could simply update the system by removing $\alpha_{l-1}$ from $\bar \alpha$ and the corresponding rows from $\bar \vy$ and $\bar \mPhi \bar \mPsi$. But if the system in \eqref{eq:y=PhiPsia_bar} has overlapping rows, where the rows are coupled with more than one set of variables, instead of removing the rows, we remove the columns. Thus, removing $\alpha_{l-1}$ is equivalent to removing the first $l_{l-1}$ coefficients from the vector $\bar \alpha$, removing the first $l_{l-1}$ columns from the matrix $\bar \mPhi \bar \mPsi$ in \eqref{eq:y=PhiPsia_bar}, and modifying the measurement vector $\bar \vy$ accordingly.
To do this we divide $\bar \vx$ into two parts as
\begin{equation}\label{eq:x_divide_2}
\bar \vx = \bar \mPsi \bar \alpha = \amatrix{\breve \mPsi & \tilde \mPsi}\amatrix{\breve \alpha \\ \tilde \alpha}= \breve \mPsi \breve \alpha + \tilde \mPsi \tilde \alpha,
\end{equation}
where we divided $\bar \mPsi$ into two matrices $\breve \mPsi$ and $\tilde \mPsi$ and $\bar \alpha$ into the corresponding vectors $\breve \alpha$ and $\tilde \alpha$.
An example of such a decomposition is depicted in Fig.~\ref{fig:yt=PhiPsiat}.
To remove $\alpha_{l-1}$ from the system in \eqref{eq:y=PhiPsia_bar}, we modify $\bar \vy$ as follows.
Since we only have an estimate of $\alpha_{l-1}$, which we denote as $\widehat \alpha_{l-1}$, we remove its expected contribution from the system by modifying $\bar \vy$ as
\begin{equation}\label{eq:yt_LOT_1}
\tilde \vy \deq \bar \vy - \bar \mPhi \breve \mPsi \breve \alpha,
\end{equation}
where we use $\breve \mPsi$ to denote the first $l_{l-1}$ columns in $\bar \mPsi$, which contains a part of $\Psi_{l-1}$, and $\breve \alpha$ to denote $\widehat \alpha_{l-1}$.
We write the resultant, modified form of the system in \eqref{eq:y=PhiPsia_bar} as
\begin{equation}\label{eq:y=PhiPsia_tilde}
\tilde \vy = \bar \mPhi \tilde \mPsi \tilde \alpha + \tilde \ve,
\end{equation}
where $\tilde \ve$ denotes combined error in the system and $\tilde \alpha$ denotes the unknown vector of coefficients that we estimate by solving a weighted $\ell_1$-norm minimization problem.

%---------%
% Figures
%---------%
%

\newif\iffullfigure
% \fullfiguretrue
% \fullfigurefalse

\subsubsection{Recovery problem}\label{sec:recovery_LOT}
To estimate $\tilde \alpha$ from the system in \eqref{eq:y=PhiPsia_tilde}, we solve the following optimization problem:
\begin{equation}\label{eq:BPDN_LOT}
\underset{\alpha}{\text{minimize}}\; \|\mW \alpha\|_1 + \frac{1}{2} \|\bar \mPhi \tilde \mPsi \alpha - \tilde \vy\|_2^2,
\end{equation}
where $\mW$ is a diagonal matrix that consists of positive weights.
We select the weights using prior knowledge about the estimate of $\tilde \alpha$ from the previous streaming iteration. Let us denote $\widehat {\alpha}$ as our prior estimate of $\tilde \alpha$. Since there is a significant overlap between the active intervals at the present and the previous iterations, we expect $\widehat \alpha$ to be very close to the solution of \eqref{eq:BPDN_LOT}.
We compute $i^\text{th}$ diagonal entry in $\mW$ as
\begin{equation}\label{eq:update_wt}
    \vw_i \gets \frac{\tau}{\beta|\widehat {\alpha_i}| + 1},
\end{equation}
where $\tau > 0$ and $\beta>>1$ are two parameters that can be used to tune the weights according to the problem.
Instead of solving \eqref{eq:BPDN_LOT} from scratch, we can speed up the recovery process by providing $\widehat {\alpha}$ as a warm-start (initialization) vector to an appropriate solver.

We compute $\widehat \alpha$ using the signal estimate from the previous streaming iteration and the available set of measurements.
Since we have an estimate of $\bar \vx$ for the part of $\Pi$ that is common between the current and the previous iteration, our main task is to predict the signal values that are new to the system.
Let us denote the available signal estimate for $\bar \vx$ as $\widehat \vx$. We can assign values to the new locations in $\widehat \vx$ using zero padding, periodic extension, or symmetric extension, and compute $\widehat \alpha$ from $\widehat \vx = \bar \mPsi \widehat \alpha$.
In our experiments, first we update $\widehat \vx$ by symmetric signal extension onto new locations and identify a candidate support for the new coefficients in $\widehat \alpha$; then we calculate magnitudes of the new coefficients by solving a least-squares problem, restricted to the chosen support, using the corresponding measurements in \eqref{eq:y=PhiPsia_bar}; and finally we truncate extremely small values of the least-squares solution and update $\widehat \alpha$.

%---------
% KF
%---------
%
\subsection{Streaming signals with linear dynamic model}\label{sec:streaming_KF}
% Measurements: \bar \vy = \bar \mPhi \bar \vx + \bar \ve
% Representation: \bar \vx = \bar \mPsi \bar \alpha
% linear dynamical model: x_k+1 = F_k x_k + f_k --> \bar \mF \bar \vx = \bar \vf
\subsubsection{System model}\label{sec:model_KF}
To include a linear dynamic model for the time-varying signal into the system, we append equations for the dynamic model to the system in \eqref{eq:y=Phix_bar}. Consider the dynamic model in \eqref{eq:LDSx}: $x_{t+1} = F_t x_t + f_t$, where the consecutive $x_t$ are non-overlapping.
At every streaming iteration, we describe a combined system of prediction equations for the $x_t$ that belong to $\bar \vx$ as follows. Suppose $\bar \vx$ contains $x_l,\ldots,x_r$, and an estimate of $x_{l-1}$, which was removed from $\bar \vx$ and committed to the output, is given as $\widehat x_{l-1}$. We rearrange the equations in \eqref{eq:LDSx} for $t=l$ as $-F_{l-1} \widehat x_{l-1} = -x_l+f_{l-1}$ and for the rest of $t$ as $0 = F_t x_t - x_{t+1} + f_t$. We stack these equations on top of one another to write the following compact form:
\begin{equation}\label{eq:KF_pred_bar}
\bar \vq = \bar \mF \bar \vx  + \bar \vf.
\end{equation}
$\bar \mF$ denotes a banded matrix that consists of negative identity matrices in the main diagonal and $F_l,\ldots,F_r$ below the diagonal; $\bar \vf$ denotes the combined prediction error; $\bar \vq$ denotes a vector that contains $-F_{l-1}\widehat x_{l-1}$ followed by zeros.

Combining the systems in \eqref{eq:y=Phix_bar} and \eqref{eq:KF_pred_bar} with the sparse representation ($\bar \vx = \bar \mPsi \bar \alpha$), we write the modified system over active interval $\Pi$ as
\begin{equation}\label{eq:KF_system}
\amatrix{\bar \vy \\ \bar \vq}  = \amatrix{\bar \mPhi  \\ \bar \mF} \bar \mPsi \bar \alpha + \amatrix{\bar \ve\\\bar \vf}.
\end{equation}
As we discussed in Sec.~\ref{sec:model_LOT} that using \eqref{eq:x_divide_2} we can remove those components of $\bar \alpha$ from the system that are committed to the output. Following the same procedure, if we want to remove a vector $\alpha_{l-1}$ that belongs to $\bar \alpha$ from the system, we modify the system in \eqref{eq:KF_system} as
\begin{equation}\label{eq:KF_system2}
\amatrix{\tilde \vy \\ \tilde \vq}  \deq \amatrix{\bar \vy \\ \bar \vq}- \amatrix{\bar \mPhi  \\ \bar \mF} \breve \mPsi \breve \alpha,
\end{equation}
where $\breve \alpha$ denotes $\widehat \alpha_{l-1}$ that is the estimate of $\alpha_{l-1}$ and $\breve \mPsi$ denotes the columns in $\bar \mPsi$ that correspond to the locations of $\alpha_{l-1}$ in $\bar \alpha$.
We represent the modified form of the system as
\begin{equation}\label{eq:KF_system_mod}
\amatrix{\tilde \vy \\ \tilde \vq}  = \amatrix{\bar \mPhi  \\ \bar \mF} \tilde \mPsi \tilde \alpha + \amatrix{\tilde \ve \\ \tilde \vf}.
\end{equation}

\subsubsection{Recovery problem}\label{sec:recovery_KF}
To estimate $\tilde \alpha$ from the system in \eqref{eq:KF_system_mod}, we solve the following optimization problem:
\begin{equation}\label{eq:BPDN_KF}
\underset{\alpha}{\text{minimize}}\; \|\mW \alpha\|_1 + \frac{1}{2} \|\bar \mPhi \tilde \mPsi \alpha - \tilde \vy\|_2^2 + \frac{\lambda}{2} \|\bar \mF \tilde \mPsi \alpha - \tilde \vq\|_2^2,
\end{equation}
where $\mW$ is a diagonal matrix that consists of positive weights and $\lambda>0$ is a regularization parameter that controls the effect of the dynamic model on the solution.
We select the weights using a prior estimate of $\tilde \alpha$, which we denote as $\widehat \alpha$. Estimate of a significant portion of $\widehat \alpha$ is known from the previous streaming iteration, and only a small portion is new to the system. We predict the incoming portion of the signal, say $x_r$, using the prediction matrix in \eqref{eq:LDSx} and the signal estimate from the previous iteration as $\widehat x_{r|r-1} \deq F_{r-1} \widehat x_{r-1}$.
We update the coefficients in $\widehat \alpha$ accordingly and set very small coefficients in $\widehat \alpha$ to zero. % xh(abs(xh)<tau/sqrt(log(P*N))) = 0;
We compute $\mW$ according to \eqref{eq:update_wt} using $\widehat{\alpha}$.
Similarly, instead of solving \eqref{eq:BPDN_KF} from scratch, we use $\widehat {\alpha}$ as a warm-start. In the next section we describe a homotopy algorithm for such a warm-start update.

% l1 homotopy
\section{$\ell_1$-homotopy: a unified homotopy algorithm}\label{sec:l1homotopy}
In this section, we present a general homotopy algorithm that we will use to dynamically update the solutions of the $\ell_1$ problems, described in \eqref{eq:BPDN_LOT} and \eqref{eq:BPDN_KF}, for the recovery of streaming, time-varying signals.
%Our proposed algorithm accepts an arbitrary warm-start, initial vector and updates

Suppose $\vy$ is a vector that obeys the following linear model: $ \vy = \mA \bar \vx+\ve,$ where $\bar \vx$ is a sparse, unknown signal of interest, $\mA$ is an $M\times N$ system matrix, and $\ve$ is a noise vector. We want to solve the following $\ell_1$-norm minimization problem to recover $\bar \vx$:
\begin{equation}\label{eq:l1homotopy_BPDN}
\underset{\vx}{\text{minimize}}\; \|\mW \vx\|_1 + \frac{1}{2} \|\mA \vx - \vy\|_2^2,
\end{equation}
where $\mW$ is a diagonal matrix that contains positive weights $\vw$ on its diagonal.
Instead of solving \eqref{eq:l1homotopy_BPDN} from scratch, we want to expedite the process by using some prior knowledge about the solution of \eqref{eq:l1homotopy_BPDN}. In this regard, we assume that we have a sparse vector, $\widehat \vx$, with support\footnote{We use the terms support and active set interchangeably for the index set of nonzero coefficients.} $\widehat \Gamma$ and sign sequence $\widehat \vz$ that is close to the original solution of \eqref{eq:l1homotopy_BPDN}.
The homotopy algorithm we present can be initialized with an arbitrary vector $\widehat \vx$, given the corresponding matrix $\mA_{\widehat \Gamma}^T\mA_{\widehat\Gamma}$ is invertible; however, the update will be quick if $\widehat \vx$ is close to the final solution.

Homotopy methods provide a general framework to solve an optimization program by continuously transforming it into a related problem for which the solution is either available or easy to compute. Starting from an available solution, a series of simple problems are solved along the so-called \emph{homotopy path} towards the final solution of the original problem \cite{Vanderbei_book_LPextension,OsbornePresnell_2000_NewApproachLasso,Efron_2004_LARS}. The progression along the homotopy path is controlled by the \emph{homotopy parameter}, which usually varies between 0 and 1, corresponding to the two end points of the homotopy path.

We build the homotopy formulation for \eqref{eq:l1homotopy_BPDN}, using $\epsilon\in [0,1]$ as the homotopy parameter, as follows. We treat the given warm-start vector $\widehat \vx$ as a starting point and solve the following optimization problem:
\begin{equation}\label{eq:l1homotopy}
\underset{\vx}{\text{minimize}}\; \|\mW \vx\|_1 + \frac{1}{2} \|\mA \vx - \vy\|_2^2 + (1-\epsilon)\vu^T\vx
\end{equation}
by changing $\epsilon$ from 0 to 1. We define $\vu$ as
\begin{equation}\label{eq:l1homotopy_u}
\vu \deq -\mW\:\widehat \vz-\mA^T(\mA \widehat \vx-\vy),
\end{equation}
where $\widehat \vz$ can be any vector that is defined as $\sign{\widehat \vx}$ on $\widehat \Gamma$ and strictly smaller than one elsewhere. Using the definition of $\vu$ in \eqref{eq:l1homotopy_u} and the conditions in \eqref{eq:l1homotopy_opts} below, we can establish that $\widehat \vx$ is the optimal solution of \eqref{eq:l1homotopy} at $\epsilon = 0$.
As $\epsilon$ changes from 0 to 1, the optimization problem in \eqref{eq:l1homotopy} gradually transforms into the one in \eqref{eq:l1homotopy_BPDN}, and the solution of \eqref{eq:l1homotopy} follows a piece-wise linear homotopy path from $\widehat \vx$ toward the solution of \eqref{eq:l1homotopy_BPDN}.
To demonstrate these facts and derive the homotopy algorithm, we analyze the optimality conditions for \eqref{eq:l1homotopy} below.

The optimality conditions for \eqref{eq:l1homotopy} can be derived by setting the subdifferential of its objective function to zero \cite{Boyd_book_ConvexOptimization,Bertsekas_1999_NonlinearProg_book}.
We can describe the conditions that a vector $\vx^*$ needs to satisfy to be an optimal solution as
\begin{equation}
\mW \vg + \mA^T(\mA\vx^*-\vy) + (1-\epsilon) \vu= 0, \;\; \|\vg\|_\infty \le 1, \;\; \vg^T\vx^* = \|\vx^*\|_1,
\end{equation}
where $\vg=\partial \|\vx^*\|_1$ denotes the subdifferential of the $\ell_1$ norm of $\vx^*$~\cite{rockafellar1997convex,Donoho_2006_FastLl1}.
This implies that for any given value of $\epsilon \in [0,~1]$, the solution $\vx^*$ for \eqref{eq:l1homotopy} must satisfy the following optimality conditions:
\begin{subequations}\label{eq:l1homotopy_opts}
\begin{align}
&&\va_i^T(\mA\vx^*-\vy) + (1-\epsilon)\vu_i &= -\vw_i \vz_i \quad &\text{for all } i \in \Gamma && \label{eq:l1homotopy_opts_a} \\
&&|\va_i^T(\mA\vx^*-\vy) + (1-\epsilon)\vu_i| &\le \vw_i \quad &\text{for all } i \in \Gamma^c,  \label{eq:l1homotopy_opts_b} &&
\end{align}
\end{subequations}
where $\va_i$ denotes $i^\text{th}$ column of $\mA$, $\Gamma$ is the support of $\vx^*$, and $\vz$ is its sign sequence.
The optimality conditions in \eqref{eq:l1homotopy_opts} can be viewed as $N$ constraints on $\va_i^T(\mA\vx-\vy) + (1-\epsilon)\vu_i$ that the solution $\vx^*$ needs to satisfy with equality (in terms of the magnitude and the sign) on the active set $\Gamma$ and strict inequality (in terms of the magnitude) elsewhere. The only exception is at the critical values of $\epsilon$ when the support changes and the constraint on the incoming or outgoing index holds with equality.
Equivalently, the locations of the active constraints in \eqref{eq:l1homotopy_opts} determine the support of $\vx^*$, $\Gamma$, and their signs determine the signs of $\vx^*$, $\vz$, which in our formulation are opposite to the signs of the active constraints.
Note that, the definition of $\vu$ in \eqref{eq:l1homotopy_u} ensures that $\widehat \vx$ satisfies the optimality conditions in \eqref{eq:l1homotopy_opts} at $\epsilon = 0$; hence, it is a valid initial solution.
It is also evident from \eqref{eq:l1homotopy_opts_a} that at any value of $\epsilon$ the solution $\vx^*$ is completely described by the support $\Gamma$ and the sign sequence $\vz$ (assuming that $(\mA_\Gamma^T\mA_\Gamma)^{-1}$ exists).
The support changes only at certain critical values of $\epsilon$, when either a new element enters the support or an existing nonzero element shrinks to zero. These critical values of $\epsilon$ are easy to calculate at any point along the homotopy path, and the entire path (parameterized by $\epsilon$) can be traced in a sequence of computationally inexpensive homotopy steps.

For every homotopy step we jump from one critical value of $\epsilon$ to the next while updating the support of the solution, until $\epsilon$ is equal to 1. As we increase $\epsilon$ by a small value $\delta$, the solution moves in a direction $\delx$, which to maintain optimality must obey
\begin{subequations}\label{eq:l1homotopy_update}
\begin{align}
&&\va_i^T(\mA\vx^*-\vy) + (1-\epsilon)\vu_i + \delta (\va_i^T\mA\delx - \vu_i) &=  -\vw_i \vz_i  \quad & \text{for all } i \in \Gamma && \label{eq:l1homotopy_update_a} \\
&&|\underbrace{\va_i^T(\mA\vx^*-\vy) + (1-\epsilon)\vu_i}_{\vp_i} + \delta \underbrace{(\va_i^T\mA\delx - \vu_i)}_{\vd_i}| &\le  \vw_i \quad &\text{for all } i \in \Gamma^c, &&  \label{eq:l1homotopy_update_b}
\end{align}
\end{subequations}
% \begin{subequations}\label{eq:l1homotopy_update}
%\begin{align}
%&&\va_i^T(\mA\vx^*-\vy) + (1-\epsilon)\vu_i + \delta (\va_i^T\mA\delx - \vu_i) &=  -\vw_i \vz_i && \notag \\
%&&& \text{for all } i \in \Gamma && \label{eq:l1homotopy_update_a} \\
%&&|\underbrace{\va_i^T(\mA\vx^*-\vy) + (1-\epsilon)\vu_i}_{\vp_i} + \delta \underbrace{(\va_i^T\mA\delx - \vu_i)}_{\vd_i}| &\le  \vw_i && \notag \\ &&& \text{for all } i \in \Gamma^c, &&  \label{eq:l1homotopy_update_b}
%\end{align}
%\end{subequations}
The update direction that keeps the solution optimal as we change $\delta$ can be written as
\begin{equation}\label{eq:l1homotopy_delx}
\delx = \begin{cases} (\mA^T_\Gamma \mA_\Gamma)^{-1} \vu_\Gamma & \text{on } \Gamma \\
0 & \text{otherwise}.
\end{cases}
\end{equation}
We can move in direction $\delx$ until either one of the constraints in \eqref{eq:l1homotopy_update_b} is violated, indicating that we must add an element to the support $\Gamma$, or one of the nonzero elements in $\vx^*$ shrinks to zero, indicating that we must remove an element from $\Gamma$.
The smallest step-size that causes one of these changes in the support can be easily computed as $\delta^* = \min(\delta^+,\delta^-)$, where\footnote{To include the positivity constraint in the optimization problem \eqref{eq:l1homotopy_BPDN}, we initialize the homotopy with a non-negative (feasible) warm-start vector and define $\delta^+= \min_{i\in \Gamma^c} \left(\frac{-\vw_i-\vp_i}{\vd_i}\right)_+$ \cite{Asif_2013_phdthesis}.}
\begin{subequations}\label{eq:l1homotopy_delta}
\begin{flalign}
&& \delta^+ &= \min_{i\in \Gamma^c} \left(\frac{\vw_i-\vp_i}{\vd_i}, \frac{-\vw_i-\vp_i}{\vd_i}\right)_+&&\\
\text{and}&& \delta^- &= \min_{i\in \Gamma} \left(\frac{-\vx^*_i}{\delx_i}\right)_+,&&
\end{flalign}
\end{subequations}
and $\min(\cdot)_+$ means that the minimum is taken over only positive arguments.
$\delta^+$ is the smallest step-size that causes an inactive constraint to become active at index $\gamma^+$, indicating that $\gamma^+$ should enter the support and $\vz_{\gamma^+}$ should be opposite to the sign of the active constraint at $\gamma^+$, and $\delta^-$ is the smallest step-size that shrinks an existing element at index $\gamma^-$ to zero, indicating that $\gamma^-$ should leave the support.
The new critical value of $\epsilon$ becomes $\epsilon+\delta^*$ and the new signal estimate $\vx^*$ becomes $\vx^* + \delta^* \delx$, and its support and sign sequence are updated accordingly.  If $\gamma^+$ is added to the support, at the next iteration we check whether the value of $\delx_{\gamma^+}$ has the same sign as $\vz_{\gamma^+}$; if the signs mismatch, we immediately remove $\gamma^+$ from the support and recompute the update direction $\delx$.

At every step along the homotopy path, we compute the update direction, the step-size, and the consequent one-element change in the support. We repeat this procedure until $\epsilon$ is equal to 1. The pseudocode outlining the homotopy procedure is presented in Algorithm~\ref{alg:l1homotopy}.

%---------------------------------
% Pseudocode
%---------------------------------
\begin{algorithm}[t]
  \caption{$\ell_1$-\textsc{Homotopy}
    \label{alg:l1homotopy}}
  \begin{algorithmic}[1]
    % \Require{$\mA$, $\vy$, $\vw$, and $\widehat \vx$ (along with $\widehat \Gamma$, $\widehat \vz$, $\vu$, and inverse or factors of $\mA_{\widehat \Gamma}^T\mA_{\widehat \Gamma}$})
    \Require{$\mA$, $\vy$, $\mW$, $\widehat \vx$, and $\vu$ (optional: inverse or decomposition factors of  $\mA_{\widehat \Gamma}^T\mA_{\widehat \Gamma}$)}
    \Ensure{$\vx^*$}
    \Statex
    \State {\bf Initialize: }$\epsilon = 0$, $\vx^* \gets \widehat\vx$
    \Repeat
    \State Compute $\delx$ in \eqref{eq:l1homotopy_delx}   \Comment{Update direction}
    \State Compute $\vp$ and $\vd$ in \eqref{eq:l1homotopy_update_b}
    \State Compute $\delta^* = \min(\delta^+,\delta^-)$ in \eqref{eq:l1homotopy_delta} \Comment{Step size}
    \If{$\epsilon + \delta^* > 1$}
    \State $\delta^* \gets 1-\epsilon$ \Comment{Last iteration}
    \State $\vx^* \gets \vx^* +\delta^* \delx$ \Comment{Final solution}
    \State \texttt{break}
    \EndIf
    \State $\vx^* \gets \vx^* +\delta^* \delx$ \Comment{Update the solution}
    \State $\epsilon \gets \epsilon +\delta^*$ \Comment{Update the homotopy parameter}
    \If{$\delta^* = \delta^-$}
        \State $\Gamma \gets \Gamma \backslash \gamma^-$ \Comment{Remove an element from the support}
    \Else
        \State $\Gamma \gets \Gamma \cup \gamma^+$ \Comment{Add a new element to the support}
    \EndIf
    \Until{$\epsilon = 1$}
  \end{algorithmic}
\end{algorithm}

The main computational cost of every homotopy step comes from computing $\delx$ by solving an $S\times S$ system of equations in \eqref{eq:l1homotopy_delx} (where $S$ denotes the size of $\Gamma$) and from computing the vector $\vd$ in \eqref{eq:l1homotopy_opts_b} that is used to compute the step-size $\delta$ in \eqref{eq:l1homotopy_delta}. Since we know the values of $\vd$ on $\Gamma$ by construction and $\delx$ is nonzero only on $\Gamma$, the cost for computing $\vd$ is same as one application of an $M\times N$ matrix. Moreover, since $\Gamma$ changes by a single element at every homotopy step, instead of solving the linear system in \eqref{eq:l1homotopy_delx} from scratch, we can efficiently compute $\delx$ using a rank-one update at every step:
\renewcommand{\labelitemi}{{\tiny$\rhd$}}
\begin{itemize}
\item {\it Update matrix inverse:}
    We can derive a rank-one updating scheme by using matrix inversion lemma to explicitly update the inverse matrix $(\mA_\Gamma^T\mA_\Gamma)^{-1}$, which has an equivalent cost of performing one matrix-vector product with an
    $M\times S$ and an $S \times S$ matrix each and adding a rank-one matrix to $(\mA^T_\Gamma\mA_\Gamma)^{-1}$. The update direction $\delx$ can be recursively computed with a vector addition. The total cost for rank-one update is approximately $MS+2S^2$ flops.
\item {\it Update matrix factorization:}
    Updating the inverse of matrix often suffers from numerical stability issues, especially when $S$ becomes closer to $M$ (i.e, the number of columns in $\mA_\Gamma$ becomes closer to the number of rows). In general, a more stable approach is to update a Cholesky factorization of $\mA_\Gamma^T\mA_\Gamma$ (or a QR factorization of $\mA_\Gamma$) as the support changes~\cite[Chapter~12]{Golub_1996_MatrixComputation}, \cite[Chapter~3]{Bjorck_1996_NumericalLS_book}. The computational cost for updating Cholesky factors and $\delx$ involves nearly $MS+3S^2$ flops.
\end{itemize}
As such, the computational cost of a homotopy step is close to the cost of one application of each $\mA$ and $\mA^T$ (that is, close to $MN+MS+3S^2+O(N)$ flops, assuming $S$ elements in the support).
If the inverse or factors of $\mA_{\widehat \Gamma}^T\mA_{\widehat \Gamma}$ are not readily available during initialization, then updating or computing that would incur an additional one-time cost.

The homotopy method described above is a versatile algorithm that can dynamically update the solution of $\ell_1$ problem in \eqref{eq:l1homotopy_BPDN} for various changes. For instance, adding or removing sequential measurements, updating solution for a time-varying signal, updating the weights, or making arbitrary changes in the system matrix. Almost all these variations appear in the $\ell_1$ problems for the recovery of streaming signals described in \eqref{eq:BPDN_LOT} and \eqref{eq:BPDN_KF}.
Similar to the homotopy formulation in \eqref{eq:l1homotopy}, we use the given $\widehat\alpha$ as a warm-start vector and solve  \eqref{eq:BPDN_LOT} at every streaming iteration using the following homotopy program:
\begin{equation}\label{eq:l1homotopy_LOT}
\underset{\alpha}{\text{minimize}}\; \|\mW \alpha\|_1 + \frac{1}{2} \|\bar \mPhi \tilde \mPsi \alpha - \tilde\vy\|_2^2 + (1-\epsilon)\vu^T\alpha,
\end{equation}
by changing $\epsilon$ from 0 to 1. To solve \eqref{eq:l1homotopy_LOT}, we provide the following parameters to Algorithm~\ref{alg:l1homotopy}: the warm-start vector  $\widehat \alpha$, the system matrix $\mA \gets \bar \mPhi\tilde \mPsi$, and the measurement vector $\vy\gets \tilde\vy$. We define $\vu$ as
\begin{equation}\label{eq:l1homotopy_u_LOT}
\vu \deq -\mW\:\widehat \vz-(\bar \mPhi \tilde \mPsi)^T(\bar \mPhi \tilde \mPsi \widehat {\alpha}-\tilde \vy),
\end{equation}
where $\widehat \vz$ can be any vector that is defined as $\sign{\widehat \alpha}$ on the support (nonzero indices) of $\widehat {\alpha}$ and strictly smaller than one elsewhere.
Similarly, to solve \eqref{eq:BPDN_KF}, we use the given warm-start vector $\widehat \alpha$ and solve the following homotopy formulation:
\begin{equation}\label{eq:l1homotopy_KF}
\underset{\alpha}{\text{minimize}}\; \|\mW \alpha\|_1 + \frac{1}{2} \|\bar \mPhi \tilde \mPsi \alpha - \tilde\vy\|_2^2  + \frac{\lambda}{2} \|\bar \mF \tilde \mPsi \alpha - \tilde \vq\|_2^2 + (1-\epsilon)\vu^T\alpha,
\end{equation}
by changing $\epsilon$ from 0 to 1, using system matrix
$\mA \gets [\bar \mPhi\tilde \mPsi~;~\sqrt{\lambda} \bar \mF \tilde \mPsi]$ and measurement vector
$\vy \gets [\tilde \vy~;~\sqrt{\lambda}\tilde \vq]$ in Algorithm~\ref{alg:l1homotopy}.
% $\mA  \gets \begin{bmatrix}
% \bar \mPhi\tilde \mPsi\\ \sqrt{\lambda} \bar \mF \tilde \mPsi
% \end{bmatrix}$ and measurement vector
% $\vy \gets \begin{bmatrix}
% \tilde \vy\\ \sqrt{\lambda}\tilde \vq
% \end{bmatrix}$ in Algorithm~\ref{alg:l1homotopy}.
We define $\vu$ as
\begin{equation}\label{eq:l1homotopy_u_KF}
\vu \deq -\mW\:\widehat \vz-(\bar \mPhi \tilde \mPsi)^T(\bar \mPhi \tilde \mPsi \widehat {\alpha}-\tilde \vy) -\lambda\,(\bar \mF \tilde \mPsi)^T(\bar \mF \tilde \mPsi \widehat {\alpha}-\tilde \vq),
\end{equation}
where $\widehat \vz$ is defined as before.
% where $\widehat \vz$ can be any vector that is defined on the support (nonzero indices) of $\widehat {\alpha}$ as $\sign{\widehat {\alpha}}$, and that is strictly smaller than one elsewhere.

% experiments
\section{Numerical experiments}\label{sec:exp}
We present experiments for the recovery of two types of time-varying signals from streaming, compressive measurements:
1) signals that have sparse representation in LOT bases and 2) signals that vary according to a linear dynamic model in \eqref{eq:LDSx} and have sparse representation in wavelet bases. We demonstrate the performance of our proposed recovery algorithms for these signals at different compression factors. We compare the performance of $\ell_1$-homotopy algorithm against two state-of-the-art $\ell_1$ solvers and demonstrate that $\ell_1$-homotopy requires significantly lesser computation operations and time.

\subsection{Signals with LOT representation}
\subsubsection{Experiment setup}
In these experiments, we used the following two discrete-time signals, $x[n]$, from the Wavelab toolbox~\cite{Wavelab}, that have sparse representation in LOT bases: 
1) {\tt LinChirp}, which is a critically sampled sinusoidal chirp signal and its frequency increases linearly from zero to one-half of the sampling frequency.
2) {\tt MishMash}, which is a summation of a quadratic and a linear chirp with increasing frequencies and a sinusoidal signal.
For both the signals, we generated $2^{15}$ samples and prepended them with $N=256$ zeros. Snapshots of \texttt{LinChirp} and \texttt{MishMash} and their LOT coefficients are presented in Fig.~\ref{fig:LinChirp_snapshot} and Fig.~\ref{fig:MishMash_snapshot}, respectively.
We estimated sparse LOT coefficients of these signals from streaming, compressive measurements using the system model and the recovery procedure outlined in Sec.~\ref{sec:streaming_LOT}.

We selected the parameters for compressive measurements and the signal representation as follows.
%
% \renewcommand{\labelitemi}{{\tiny$\rhd$}}
% \renewcommand{\labelitemi}{{\tiny${}$}}
% \begin{itemize}
% {\it Compressive measurements}:
To simulate streaming, compressive measurements of a given time-varying signal, $x[n]$, at a compression rate $R$, we followed the model in \eqref{eq:LDSy}: $y_t = \Phi_t x_t + e_t$. We used non-overlapping $x_t$ of length $N$ to generate a set of $M = N/R$ measurements in $y_t$. We generated entries in $\Phi_t$ independently at random as $\pm 1/\sqrt M$ with equal probability. We added Gaussian noise in the measurements by selecting every entry in $e_t$ according to $\mathcal{N}(0,\sigma^2)$ distribution. We selected the variance $\sigma^2$ such that the expected SNR with respect to the measurements $\Phi_t x_t$ becomes $35$\,dB.
%
% {\it Signal representation}:
To represent $x[n]$ using LOT bases, according to \eqref{eq:x[n]_general}, we selected the overlapping intervals, $I_p$, of the same length $N+2\eta_p=2N$, where we fixed $\eta_p = N/2$, $a_p = pN+1/2$, and $l_p = N$ for all $p\in \mathbb{Z}$. We divided $x[n]$ into overlapping intervals, $I_p$, and computed the corresponding LOT coefficients $\alpha_p$.
% \end{itemize}

At every streaming iteration, we built the system in \eqref{eq:y=PhiPsia_tilde} for $P=5$ consecutive $x_t$ in $\bar \vx$.
We updated the system in \eqref{eq:y=Phix_bar}, from the previous iteration, by shifting the active interval, removing old measurements, and adding new measurements. We computed $\tilde \vy$ in \eqref{eq:yt_LOT_1}, committed a portion of $\widehat \alpha$ to the output.
The combined system in \eqref{eq:y=PhiPsia_tilde}, corresponding to the unknown vector $\bar \vx$ of length $PN$, thus, consists of a measurement vector $\tilde \vy$ of length $PM$, a block diagonal $PM\times PN$ measurement matrix $\bar \mPhi$, a $PN \times PN$ LOT representation matrix $\tilde \mPsi$ in which adjacent pairs of columns overlap in $N$ rows, the unknown LOT coefficient vector $\tilde \alpha$ of length $PN$, and a noise vector $\tilde \ve$. An example of such a system in depicted in Fig.~\ref{fig:activesystem}.
We predicted the new coefficients in $\widehat \alpha$, updated the weights $\mW$, and solved \eqref{eq:BPDN_LOT} using $\widehat \alpha$ as a warm-start.
We updated the weights according to \eqref{eq:update_wt} using $\beta = M\frac{\|\widehat \alpha\|_2^2}{\|\widehat \alpha\|_1^2}$ and $\tau = \max\{10^{-2}\|\mA^T \vy\|_\infty,\sigma \sqrt{\log (PN)}\}$, where $\mA$ and $\vy$ denote the system matrix and the measurement vector in \eqref{eq:BPDN_LOT}, respectively, and $\sigma$ denotes the standard deviation of the measurement noise.
For the first streaming iteration, we initialized $\alpha$ as zero and solved \eqref{eq:BPDN_LOT} as an iterative reweighted $\ell_1$ problem, starting with $\mW = \tau$, using five reweighting iterations~\cite{Candes_rwtL1_2008,AR_2012_rwtL1}.

We solved \eqref{eq:BPDN_LOT} using our proposed $\ell_1$-homotopy algorithm and two state-of-the-art $\ell_1$ solvers: YALL1~\cite{yang-2011-yall1} and SpaRSA~\cite{Wright-2009-sparsa}, with identical initialization (warm-start) and weight selection procedure. Further description of these algorithms is as follows. \\
%
% \begin{enumerate}[i.]
{\it $\ell_1$-homotopy\footnote{\label{fn:L1-homotopy}$\ell_1$-homotopy code: http://users.ece.gatech.edu/$\sim$sasif/homotopy. Additional experimental results are also available on the same page.}
%(also at https://github.com/sasif/L1-homotopy.git). Additional experiments and scripts to reproduce all the experiments in this paper can also be found on this webpage.}
}:
We solved \eqref{eq:l1homotopy_LOT} following the procedure outlined in Algorithm~\ref{alg:l1homotopy}. The main computational cost at every step of $\ell_1$-homotopy involves one matrix-vector multiplication for identifying a change in the support and a rank-one update for computing the update direction. We used the matrix inversion lemma-based scheme to perform the rank-one updates. \\
{\it YALL1\footnote{YALL1 code: http://yall1.blogs.rice.edu}}: YALL1 is a first-order algorithm that uses an alternating direction minimization method for solving various $\ell_1$ problems, see~\cite{yang-2011-yall1} for further details. We solved \eqref{eq:BPDN_LOT} using weighted-$\ell_1/\ell_2$ solver in YALL1 package by selecting the initialization vector and weights according to the procedure described in Sec.~\ref{sec:recovery_LOT}. At every streaming iteration, we used previous YALL1 solution to predict the initialization vector and the weights according to \eqref{eq:update_wt}. We fixed the tolerance parameter to $10^{-4}$ in all the experiments. The main computational cost of every step in the YALL1 solver comes from applications of $\mA$ and $\mA^T$. \\
{\it SpaRSA\footnote{SpaRSA code: http://lx.it.pt/$\sim$mtf/SpaRSA}}: SpaRSA is also a first-order method that uses a fast variant of iterative shrinkage and thresholding for solving various $\ell_1$-regularized problems, see~\cite{Wright-2009-sparsa} for further details. Similar to YALL1, we solved \eqref{eq:BPDN_LOT} using SpaRSA at every streaming iteration by selecting the initialization vector and the weights from the solution of previous iteration.
We used the SpaRSA code with default adaptive continuation procedure in the Safeguard mode using the duality gap-based termination criterion for which we fixed the tolerance parameter to $10^{-4}$ and modified the code to accommodate weights in the evaluation. The main computational cost for every step in the SpaRSA solver also involves applications of $\mA$ and $\mA^T$.
%
% \end{enumerate}

To summarize, $\ell_1$-homotopy solves homotopy formulation of \eqref{eq:BPDN_LOT}, given in \eqref{eq:l1homotopy_LOT}, while  YALL1 and SpaRSA solve \eqref{eq:BPDN_LOT} using a warm-start vector for the initialization.

We used MATLAB implementations of all the algorithms and performed all the experiments on a standard laptop computer. We used a single computational thread for all the experiments, which involved recovery of a sparse signal from a given set of streaming measurements using all the candidate algorithms. In every experiment, we recorded three quantities for each algorithm: 1) the quality of reconstructed signal in terms of signal-to-error ratio (SER) in dB, defined as
$$ \text{SER} = -10\log_{10}\frac{\|  x-\widehat x\|_2^2}{\| x\|_2^2},$$
% $$ \text{SER} = -10\log_{10}\frac{\|\bar\vx-\widehat\vx\|_2^2}{\|\bar\vx\|_2^2},$$
where $ x$ and $\widehat  x$ denote the original and the reconstructed streaming signal, respectively, 2) the number of matrix-vector products with $\mA$ and $\mA^T$, and 3) the execution time in MATLAB.

\subsection{Results}

\begin{figure*}
\centering
    \begin{subfigure}[t]{\textwidth}
    \centering
      %trim option's parameter order: left bottom right top
        % \includegraphics[trim = 0mm 0mm 0mm 0mm, clip, page = 2, scale=0.5,keepaspectratio]{results_wtBPDN}
        \includegraphics[page=2,trim = 0mm 0mm 0mm 0mm, clip, width=0.85\columnwidth,keepaspectratio]
        {results_mTypesign_sTypeLinChirp_SNR35_snapshot}
        \caption{Snapshot of \texttt{LinChirp} signal, LOT coefficients, and errors in the reconstruction. \textbf{Top left:} Signal $x[n]$ (zoomed in over first 2560 samples). \textbf{Bottom left:} LOT coefficients $\alpha_p$. \textbf{Top right:} Error in the reconstructed signal at $R=4$. \textbf{Bottom right:} Error in the reconstructed LOT coefficients}
        \label{fig:LinChirp_snapshot}
    \end{subfigure}

    \vspace{0mm}

    \begin{subfigure}[t]{\textwidth}
    \centering
      %trim option's parameter order: left bottom right top
        % \includegraphics[trim = 0mm 0mm 0mm 0mm, clip, page = 2, scale=0.5,keepaspectratio]{results_wtBPDN}
        \includegraphics[page=1,trim = 0mm 0mm 0mm 0mm, clip, width=1\columnwidth,keepaspectratio]{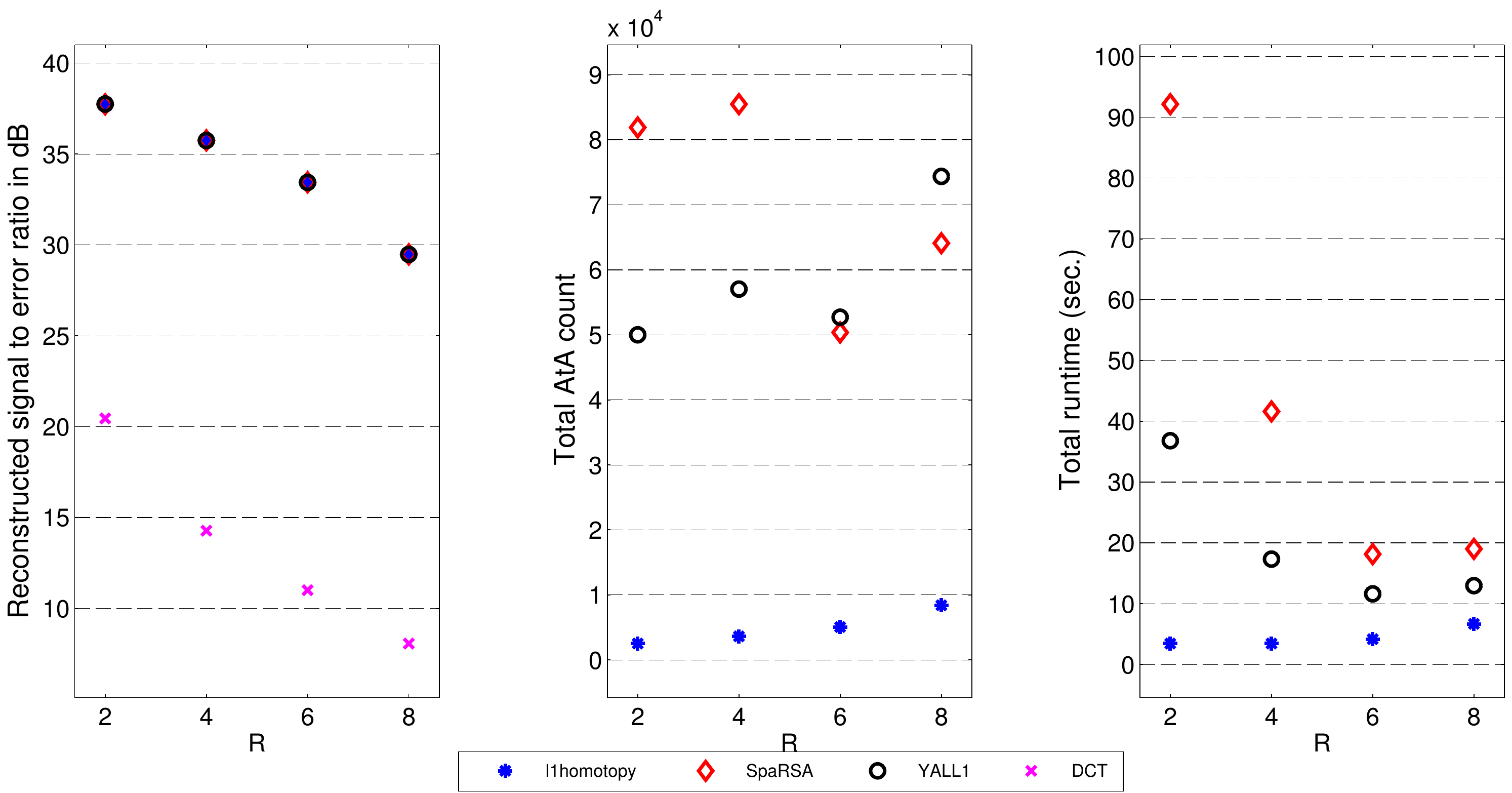}
        \caption{Results for the recovery of \texttt{LinChirp} signal from random, compressive measurements in the presence of 35dB noise. \textbf{Left:} SER at different $R$. \textbf{Middle:} Approximate count of matrix-vector multiplications. \textbf{Right:} Matlab execution time in seconds.}
        \label{fig:LinChirp_results}
    \end{subfigure}
    % \caption{\texttt{LinChirp} experiments.}\label{fig:LinChirp}
    \caption{Experiments on the \texttt{LinChirp} signal reconstruction from streaming, compressed measurements using LOT representation.}\label{fig:LinChirp}
\end{figure*}

\begin{figure*}
\centering
    \begin{subfigure}[t]{\textwidth}
    \centering
      %trim option's parameter order: left bottom right top
        % \includegraphics[trim = 0mm 0mm 0mm 0mm, clip, page = 2, scale=0.5,keepaspectratio]{results_wtBPDN}
        \includegraphics[page=2,trim = 0mm 0mm 0mm 0mm, clip, width=0.85\columnwidth,keepaspectratio]{results_mTypesign_sTypeMishMash_SNR35_snapshot}
        \caption{Snapshot of \texttt{MishMash} signal, LOT coefficients, and errors in the reconstruction. \textbf{Top left:} Signal $x[n]$ (zoomed in over first 2560 samples). \textbf{Bottom left:} LOT coefficients $\alpha_p$. \textbf{Top right:} Error in the reconstructed signal at $R=4$. \textbf{Bottom right:} Error in the reconstructed LOT coefficients}
        \label{fig:MishMash_snapshot}
    \end{subfigure}

    \vspace{0mm}

    \begin{subfigure}[t]{\textwidth}
    \centering
      %trim option's parameter order: left bottom right top
        % \includegraphics[trim = 0mm 0mm 0mm 0mm, clip, page = 2, scale=0.5,keepaspectratio]{results_wtBPDN}
        \includegraphics[page=1,trim = 0mm 0mm 0mm 0mm, clip, width=1\columnwidth,keepaspectratio]{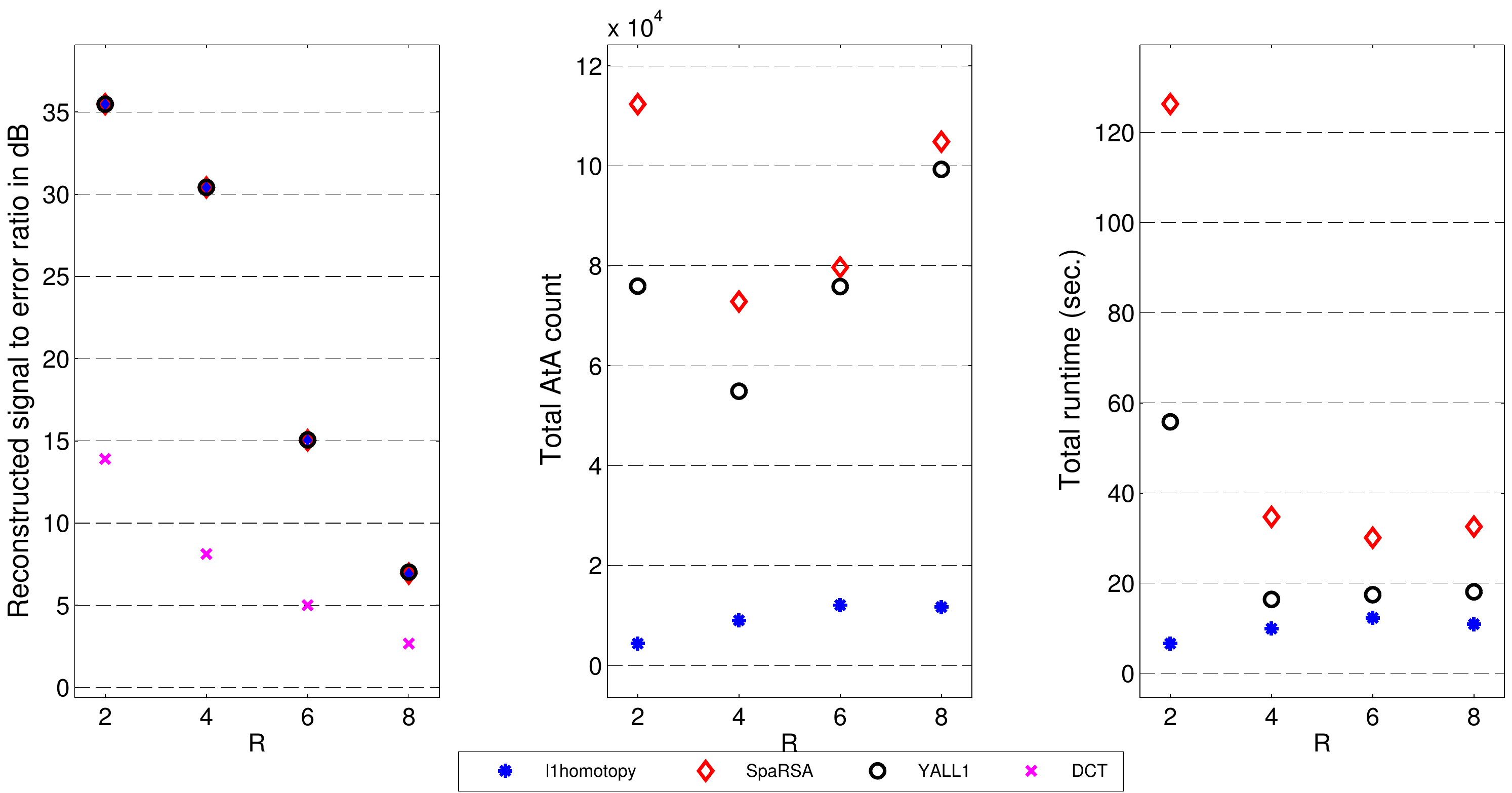}
        \caption{Results for the recovery of \texttt{MishMash} signal from random, compressive measurements in the presence of 35dB noise. \textbf{Left:} SER at different $R$. \textbf{Middle:} Approximate count of matrix-vector multiplications. \textbf{Right:} Matlab execution time in seconds.}
        \label{fig:MishMash_results}
    \end{subfigure}
    % \caption{\texttt{MishMash} experiments.}\label{fig:MishMash}
    \caption{Experiments on the \texttt{MishMash} signal reconstruction from streaming, compressed measurements using LOT representation.}\label{fig:MishMash}
\end{figure*}

We compared performances of $\ell_1$-homotopy, YALL1, and SpaRSA for the recovery of  \texttt{LinChirp} and \texttt{MishMash} signals from streaming, compressive measurements.
We performed 5 independent trials for the recovery of the streaming signal from random, streaming measurements at different values of compression factor $R$. The results, averaged over all the trials are presented in Figures~\ref{fig:LinChirp}--\ref{fig:MishMash}.

%-----------------------------------
% Figures LOT
% Experiments generated with seed 2013 in sign
% l1-homotopy : mil
% SpaRSA safeguard = 1, termination criterion = 2, tol =1e-4
% YALL1 - digit = 4.
%-----------------------------------

%-----------------------------------

\renewcommand{\figwidth}{0.33\textwidth}
\newif\iffullfigure

\fullfiguretrue
\fullfigurefalse

%-----------------------------------
% Figures (LOT)
%-----------------------------------
%-----------------------------------
% LinChirp
%-----------------------------------

%-----------------------------------
% MishMash
%-----------------------------------
%
%----------------------------
% End figures
%----------------------------

Figure~\ref{fig:LinChirp} presents results for experiments with \texttt{LinChirp} signal. Figure~\ref{fig:LinChirp_snapshot} presents a snapshot of the \texttt{LinChirp} signal, its LOT coefficients, and the reconstruction error at $R=4$. Three plots in Fig.~\ref{fig:LinChirp_results} present results for the three solvers: $\ell_1$-homotopy ({\color{blue}$*$}), SpaRSA ({\color{red} $\diamond$}), and YALL1 ($\circ$).
The left plot in Fig.~\ref{fig:LinChirp_results} compares SER for the three solvers. Since all of them solve the same convex program, SERs for the reconstructed signals are almost identical. To gauge the advantage of LOT-based reconstruction over a block transform-based reconstruction, we repeated the same experiment by replacing the LOT bases with the DCT bases for signal representation (results shown as {\color{magenta} $\times$}). We can see a significant degradation (more than $20$\,dB loss in SER) in the results for the DCT-based representation as compared to the results for the LOT-based representation.
The middle plot in Fig.~\ref{fig:LinChirp_results} compares the computational cost of all the algorithms in terms of the total number of matrix-vector multiplications used in the signal reconstruction. We counted an application of each $\mA$ and $\mA^T$ as one application of $\mA^T\mA$\footnote{For the homotopy algorithms, we approximated the cost of one step as one application of $\mA^T\mA$.}. We can see that, out of the three solvers, $\ell_1$-homotopy required the least number of $\mA^T\mA$ applications in all the experiments.
The right plot in Fig.~\ref{fig:LinChirp_results} compares the MATLAB execution time for each solver.
We can see that, compared to YALL1 and SpaRSA, $\ell_1$-homotopy consumed distinctly lesser time for the reconstruction.

Figures~\ref{fig:MishMash} presents similar results for experiments with \texttt{MishMash} signal. Figure~\ref{fig:MishMash_snapshot} presents a snapshot of the \texttt{MishMash} signal, its LOT coefficients, and the reconstruction error at $R=4$. Three plots in Fig.~\ref{fig:MishMash_results} compare the performance of the three solvers. In these plots we see similar results that the reconstruction error for \eqref{eq:BPDN_LOT} using all the solvers is almost identical, but $\ell_1$-homotopy performs significantly better in terms of the computational cost and execution time.

A brief summary of the results for our experiments is as follows. We observed that the signals reconstructed using the LOT-based representation had significantly better quality compared to those reconstructed using the DCT-based signal representation. Computational cost and execution time for $\ell_1$-homotopy is significantly smaller than that for SpaRSA and YALL1.

\subsection{Linear dynamic model}
\subsubsection{Experiment setup}
In these experiments, we simulated time-varying signal $x[n]$ according to the linear dynamic model defined in \eqref{eq:LDSx}: $x_{t+1} = F_t x_t + f_t$. We generated a seed signal of length $N=256$, which we will denote as $x_0$. Starting with $x_0$, we generated a sequence of signal instances $x_t$ for $t=1,~2,\ldots$ as follows.
% such that $x_{k+1}$ is a shifted copy of $x_k$.
For each $t$, we generated $x_{t+1}$ by applying a non-integer, left-circular shift $\epsilon_t \sim \text{uniform}(0.5,~1.5)$ to $x_t$ (i.e., $x_{t+1}[n] = x_t[(n+\epsilon_t)_{\text{mod}\,N}]$, where $\epsilon_t$ is drawn uniformly, at random from interval $[0.5,~1.5]$). We computed the $x_t$ at non-integer locations using linear interpolation. To define the dynamic model, we assumed that the individual shifts ($\epsilon_t$) are unknown and only their average value is known, which is one in our experiments. Therefore, we defined $F_t$, for all $t$, as a matrix that applies left-circular shift of one, whereas $f_t$ accounts for the prediction error in the model because of the unaccounted component of the shift $\epsilon_t$.

We used the following two signals from the Wavelab toolbox~\cite{Wavelab} as $x_0$: 1) \texttt{HeaviSine}, which is a summation of a sinusoidal and a rectangular signal and 2) \texttt{Piece-Regular}, which is a piecewise smooth signal. \texttt{HeaviSine} and \texttt{Piece-Regular} signals along with examples of their shifted copies are presented in Fig.~\ref{fig:heavisine_snapshot} and Fig.~\ref{fig:pcwreg_snapshot}, respectively.
We concatenated the $x_t$ for $t=1,~2,\ldots,128$ to build the time-varying signal $x[n]$ of length $2^{15}$.
We estimated sparse wavelet coefficients of $x[n]$ from streaming, compressive measurements using the system model and the recovery procedure outlined in Sec.~\ref{sec:streaming_KF}.

We selected the compressive measurements and the signal representation as follows.
%
% \renewcommand{\labelitemi}{{\tiny$\rhd$}}
% \renewcommand{\labelitemi}{{\tiny${}$}}
% \begin{itemize}
% {\it Compressive measurements}:
We simulated streaming, compressive measurements of $x[n]$ according to \eqref{eq:LDSy}, using the same procedure as described in the previous section. For a desired compression rate $R$, we generated $y_t$ with $M = N/R$ measurements of non-overlapping $x_t$, generated entries in $\Phi_t$ as $\pm 1/\sqrt M$ with equal probability, and added Gaussian noise in the measurements such that the expected SNR becomes $35$\,dB.
%
%{\it Signal representation}:
To represent $x[n]$ according to the model in \eqref{eq:x[n]_general}, we used (block-based) Daubechies-8 orthogonal wavelets~\cite{Daubechies_book_TenLectures} with five levels of decomposition. We divided $x[n]$ into consecutive, disjoint components, $x_t$, of length $N$ and computed wavelet coefficients, $\alpha_t$, using circular convolution in the wavelet analysis filter bank.
% In contrast with the overlapping basis representation, the sampling intervals ($\Omega_k$) and the signal representation intervals $(I_k)$ are identical in these experiments.
% \end{itemize}

At every streaming iteration, we built the system in \eqref{eq:KF_system_mod} for $P=3$ consecutive $x_t$ in $\bar \vx$.
We updated the system in \eqref{eq:KF_system} by shifting the active interval, removing old measurements, and adding new measurements. We computed $\tilde \vy$ and $\tilde \vq$ in \eqref{eq:KF_system2} and committed a portion of $\widehat \alpha$ to the output.
The combined system in \eqref{eq:KF_system_mod}, corresponding to the unknown vector $\bar \vx$ of length $PN$, thus, consists of measurement vectors $\tilde \vy,\tilde\vq$ of length $PM$ and $PN$, respectively, a block diagonal $PM\times PN$ measurement matrix $\bar \mPhi$, a banded $PN\times PN$ prediction matrix $\bar \mF$, a block-diagonal $PN \times PN$ representation matrix $\tilde \mPsi$, the unknown wavelet coefficient vector $\tilde \alpha$ of length $PN$, and error vectors $\tilde \ve,\tilde\vf$.
We predicted values of the new coefficients in $\widehat \alpha$, updated the weights $\mW$, and solved \eqref{eq:BPDN_KF} using $\widehat \alpha$ as a warm-start.
We selected $\lambda = 1/2$ and updated the weights according to \eqref{eq:update_wt} using $\beta = M\frac{\|\widehat \alpha\|_2^2}{\|\widehat \alpha\|_1^2}$ and $\tau = \max\{10^{-2}\|\mA^T \vy\|_\infty,\sigma \sqrt{\log (PN)}\}$, where $\mA $ and $\vy$ denote the system matrix and the measurements in \eqref{eq:BPDN_KF}, respectively, and $\sigma$ denotes the standard deviation of the measurement noise. We truncated the values in  $\widehat \alpha$ that are smaller than $\tau/\sqrt{\log(PN)}$ to zero.
For the first streaming iteration, we initialized $\widehat x_{l-1}$ as $x_0$ and $\alpha$ as zero. We solved \eqref{eq:BPDN_KF} as an iterative reweighted $\ell_1$ problem, starting with $\mW = \tau$, using five reweighting iterations~\cite{Candes_rwtL1_2008,AR_2012_rwtL1}.

We solved \eqref{eq:BPDN_KF} using our proposed $\ell_1$-homotopy algorithm (which in fact solves \eqref{eq:l1homotopy_KF}) and SpaRSA, with identical initialization (warm-start) and weight selection procedure. Since YALL1 only works with under-determined systems, we did not use it in these experiments.
 
\subsubsection{Results}
%Description of the experiments:
%- All the representation intervals $I_k$ have same length and we used daub8 wavelets with sym=0, J = log2(N)-3
%- Sampling intervals have length
%    $N = 256$
%- We selected $P=3$ sampling intervals in one streaming interval.
%- We tested $R=2,4,6,8$
%- Average results over 5 independent trials
%- $\pm1$ measurements
%- cshift = -1
%- rshift = @(z) (rand-0.5)
%- lambda = sqrt(0.5)
%- sig_length = 2^15 (128 copies of 256 samples)
%
%- LS-Kalman = smooth
%
% solves for x_k using the prediction covariance matrix
% from all previous measurements and smoothing with P-1 future measurements
% minimize 1/2 (x_k-\widehat{x}_k|k-1)'*P_k|k-1(x_k-\widehat{x}_k|k-1)
% + \sum_{k = i,...,i+P-1} 1/2\|y_k-A_k x_k\|_2^2 + \lambda/2\|F_k x_k -x_k+1\|_2^2

\begin{figure*}
\centering
    \begin{subfigure}[t]{\textwidth}
    \centering
      %trim option's parameter order: left bottom right top
        % \includegraphics[trim = 0mm 0mm 0mm 0mm, clip, page = 2, scale=0.5,keepaspectratio]{results_wtBPDN}
        \includegraphics[page=2,trim = 0mm 0mm 0mm 0mm, clip, width=0.85\columnwidth,keepaspectratio]{results_mTypesign_sTypeheavisine_SNR35_snapshot}
        \caption{Snapshot of the original and the reconstructed signal, error in the reconstruction, and the comparison of $\ell_1$- and $\ell_2$-regularized reconstructions. \textbf{Top left:} \texttt{HeaviSine} signal $x[n]$ drawn as an image; $p^\text{th}$ column represents $x_p$; $x_1$, $x_{20}$ and $x_{60}$ are plotted on the right. \textbf{Bottom left:} Reconstructed signal at $R=4$. \textbf{Top right:} Error in the reconstructed signal. \textbf{Bottom right:} Comparison between SERs for the solution of the $\ell_1$-regularized problem in \eqref{eq:BPDN_KF} (solid-blue line, labeled L1) and the solution of the $\ell_2$-regularized (Kalman filter smoothing) problem in \eqref{eq:Kalman-smooth} (broken-black line, labeled LS).}
        \label{fig:heavisine_snapshot}
    \end{subfigure}

    \vspace{0mm}

    \begin{subfigure}[t]{\textwidth}
    \centering
      %trim option's parameter order: left bottom right top
        % \includegraphics[trim = 0mm 0mm 0mm 0mm, clip, page = 2, scale=0.5,keepaspectratio]{results_wtBPDN}
        \includegraphics[page=1,trim = 0mm 0mm 0mm 0mm, clip, width=1\columnwidth,keepaspectratio]{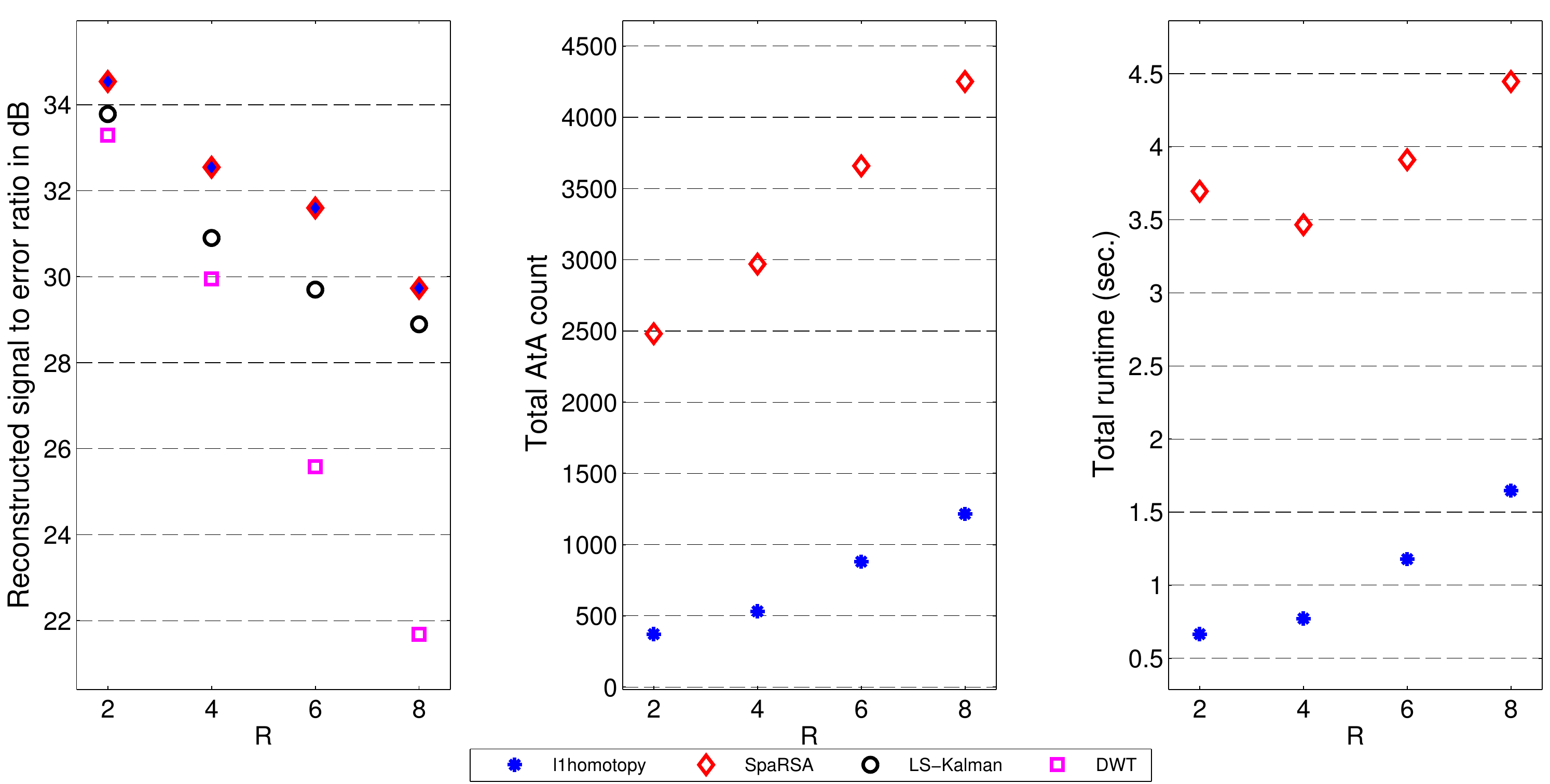}
        \caption{Results for the recovery of \texttt{HeaviSine} signal from random, compressive measurements in the presence of 35dB noise. \textbf{Left:} SER at different $R$. \textbf{Middle:} Approximate count of matrix-vector multiplications. \textbf{Right:} Matlab execution time in seconds.}
        \label{fig:heavisine_results}
    \end{subfigure}
    % \caption{\texttt{HeaviSine} experiments.}\label{fig:heavisine}
    \caption{Experiments on the time-varying \texttt{HeaviSine} signal reconstruction from streaming, compressed measurements when the signal follows a linear dynamic model.}\label{fig:heavisine}
\end{figure*}

\begin{figure*}
\centering
    \begin{subfigure}[t]{\textwidth}
    \centering
      %trim option's parameter order: left bottom right top
        % \includegraphics[trim = 0mm 0mm 0mm 0mm, clip, page = 2, scale=0.5,keepaspectratio]{results_wtBPDN}
        \includegraphics[page=2,trim = 0mm 0mm 0mm 0mm, clip, width=0.85\columnwidth,keepaspectratio]{results_mTypesign_sTypepcwreg_SNR35_snapshot}
        \caption{Snapshot of the original and the reconstructed signal, error in the reconstruction, and the comparison of $\ell_1$- and $\ell_2$-regularized reconstructions. \textbf{Top left:} \texttt{Piece-Regular} signal $x[n]$ drawn as an image; $p^\text{th}$ column represents $x_p$; $x_1$, $x_{20}$ and $x_{60}$ are plotted on the right. \textbf{Bottom left:} Reconstructed signal at $R=4$. \textbf{Top right:} Error in the reconstructed signal. \textbf{Bottom right:} Comparison between SERs for the solution of the $\ell_1$-regularized problem in \eqref{eq:BPDN_KF} (solid-blue line, labeled L1) and the solution of the $\ell_2$-regularized (Kalman filter smoothing) problem in \eqref{eq:Kalman-smooth} (broken-black line, labeled LS).}
        \label{fig:pcwreg_snapshot}
    \end{subfigure}

    \vspace{0mm}

    \begin{subfigure}[t]{\textwidth}
    \centering
      %trim option's parameter order: left bottom right top
        % \includegraphics[trim = 0mm 0mm 0mm 0mm, clip, page = 2, scale=0.5,keepaspectratio]{results_wtBPDN}
        \includegraphics[page=1,trim = 0mm 0mm 0mm 0mm, clip, width=1\columnwidth,keepaspectratio]{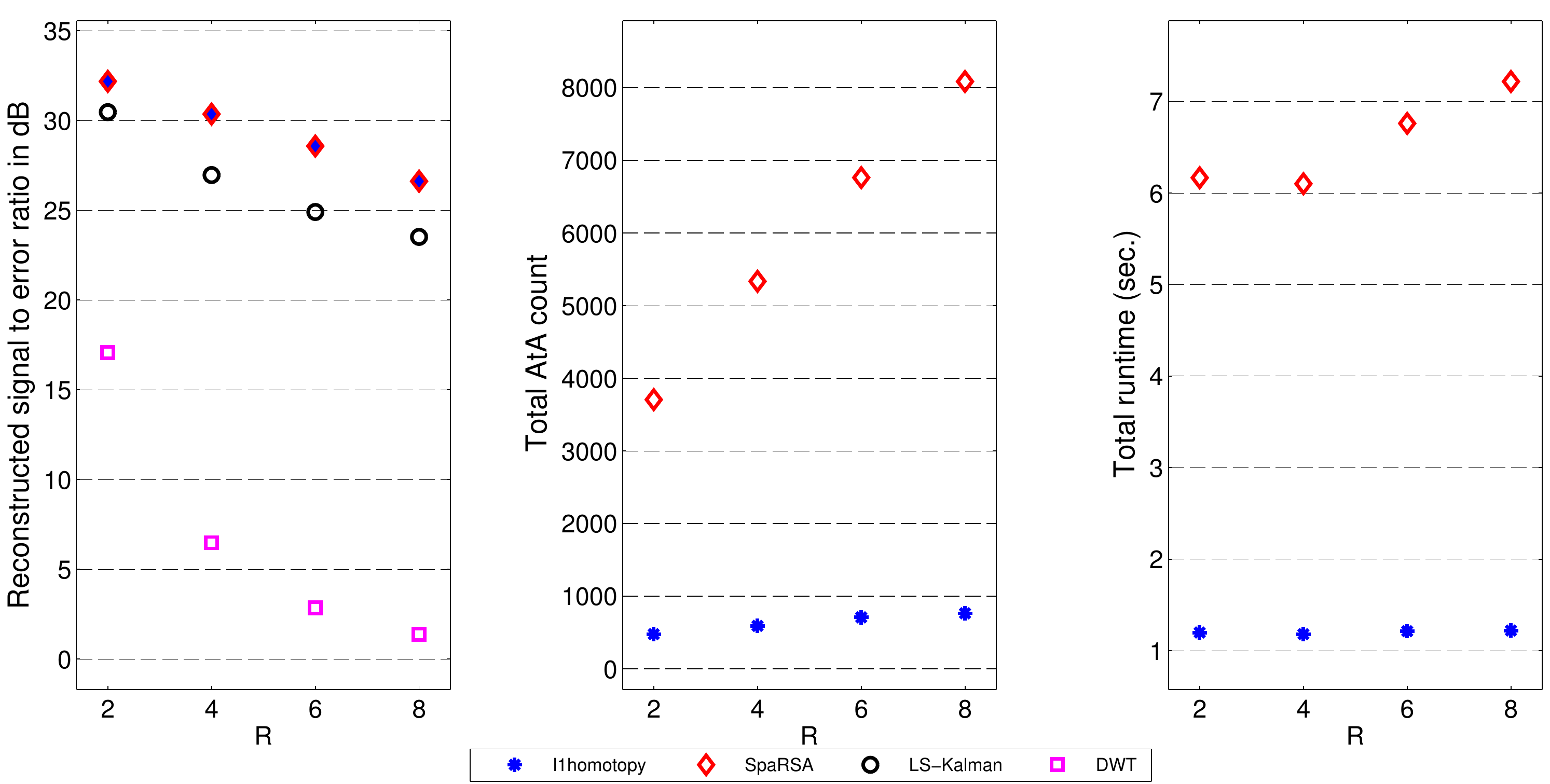}
        \caption{Results for the recovery of \texttt{Piece-Regular} signal from random, compressive measurements in the presence of 35dB noise. \textbf{Left:} SER at different $R$. \textbf{Middle:} Approximate count of matrix-vector multiplications. \textbf{Right:} Matlab execution time in seconds.}
        \label{fig:pcwreg_results}
    \end{subfigure}
   % \caption{\texttt{Piece-Regular} experiments.}\label{fig:pcwreg}
   \caption{Experiments on the time-varying \texttt{Piece-Regular} signal reconstruction from streaming, compressed measurements when the signal follows a linear dynamic model.}\label{fig:pcwreg}
\end{figure*}

We compared the performance of $\ell_1$-homotopy and SpaRSA for the recovery of \texttt{HeaviSine} and \texttt{Piece-Regular} signals from streaming, compressive measurements.
We performed 5 independent trials at different values of the compression factor $R$. In each experiment, we estimated the time-varying signal using all the algorithms, according to the procedures described above, and recorded corresponding signal-to-error ratio, number of matrix-vector products, and MATLAB runtime. The results, averaged over all the trials, are presented in Figures~\ref{fig:heavisine}--\ref{fig:pcwreg}.

%-----------------------------------
% Figures (KF)
%-----------------------------------
%-----------------------------------
% heavisine
%-----------------------------------

%-----------------------------------
% pcwreg
%-----------------------------------

%----------------------------
% End figures
%----------------------------

Figure~\ref{fig:heavisine} presents results for experiments with \texttt{HeaviSine} signal. Figure~\ref{fig:heavisine_snapshot}, top-left image presents the \texttt{HeaviSine} signal, where $p^\text{th}$ column represents $x_p$. Next to the image, we have plotted three examples for $x_1,x_{20},x_{60}$, as three colored lines. Bottom-left image is the reconstructed signal at $R=4$ along with the examples of the reconstructed $x_p$ on its right. Top-right image represents errors in the reconstruction. Bottom-right plot presents a comparison between the SER for the solution of the $\ell_1$-regularized problem in \eqref{eq:BPDN_KF} and the solution of the following $\ell_2$-regularized (Kalman filtering and smoothing) problem using the systems in \eqref{eq:y=Phix_bar} and \eqref{eq:KF_pred_bar}:
\begin{equation}\label{eq:Kalman-smooth}
\underset{\vx}{\text{minimize}}\; (x_p - \widehat{x}_{p|p-1})^T P^{-1}_{p|p-1}(x_p - \widehat{x}_{p|p-1}) + \lambda \|\tilde \mF\vx\|_2^2 + \|\bar \mPhi \vx -\bar \vy\|_2^2,
\end{equation}
where $\vx$ denotes a vector that consists of $x_p,\ldots,x_{p+P-1}$, $\tilde \mF$ denotes a submatrix of $\bar \mF$ (without its first $N$ rows), and $P_{p|p-1}$ denotes the error covariance matrix for the Kalman filter estimate $\widehat x_{p|p-1}$ given all the previous measurements \cite{Kalman-1960,Sorenson_GausstoKalman_1970}. % \cite{Kay-1993-Estimation,Kailath-2000-linear}.
% minimize 1/2 (x_k-\widehat{x}_k|k-1)'*P_k|k-1(x_k-\widehat{x}_k|k-1)
% + \sum_{k = i,...,i+P-1} 1/2\|y_k-A_k x_k\|_2^2 + \lambda/2\|F_k x_k -x_k+1\|_2^2
%
Three plots in Fig.~\ref{fig:heavisine_results} compare performance of the $\ell_1$-homotopy ({\color{blue}$*$}) and SpaRSA ({\color{red} $\diamond$}).
The left plot in Fig.~\ref{fig:heavisine_results} compares the SER for the two solvers. Since both of them solve the same convex program, SERs for the reconstructed signals are almost identical. To demonstrate the advantage of our proposed recovery framework \eqref{eq:BPDN_KF}, we present results for the solution of two related recovery problems, for identical signal representation and measurement settings:
1) Kalman filtering and smoothing problem \eqref{eq:Kalman-smooth} (labeled as LS-Kalman and plotted as $\circ$),  which does not take into account the sparsity of the signal. As the results indicate, the Kalman filter estimate is not as good as the one for the $\ell_1$-regularized problem in \eqref{eq:BPDN_KF}.
2) Weighted $\ell_1$-regularized problem in \eqref{eq:BPDN_KF} without the dynamic model, which is equivalent to solving \eqref{eq:BPDN_KF} with $\lambda = 0$, and it exploits only the sparse representation of each $x_p$ in wavelets (results labeled as DWT and plotted as {\color{magenta} $\times$}). We observed a significant degradation in the signals reconstructed without the dynamic model; the results are indeed inferior to the LS-Kalman.
The middle plot in Fig.~\ref{fig:heavisine_results} compares the computational cost of all the algorithms in terms of the total number of matrix-vector multiplications used in the signal reconstruction, and 
the right plot in Fig.~\ref{fig:heavisine_results} compares the MATLAB execution time for each solver.
We observed that $\ell_1$-homotopy consumed distinctly fewer matrix-vector multiplications and lesser computation time for the signal reconstruction.

Figures~\ref{fig:pcwreg} presents similar results for experiments with \texttt{Piece-Regular} signal. Figure~\ref{fig:pcwreg_snapshot} presents a snapshot of the \texttt{Piece-Regular} signal, its reconstruction at $R=4$ using \eqref{eq:BPDN_KF}, error in the reconstruction, and comparison between the reconstruction of \eqref{eq:BPDN_KF} and \eqref{eq:Kalman-smooth}.
Three plots in Fig.~\ref{fig:pcwreg_results} compare performance of the two solvers. In these plots we see similar results that the reconstruction error for \eqref{eq:BPDN_KF} using both $\ell_1$-homotopy and SpaRSA is almost identical, but $\ell_1$-homotopy performs significantly better in terms of computational cost and execution time.
For the DWT experiments with \texttt{Piece-Regular} signal, we solved a non-weighted version of \eqref{eq:BPDN_KF}, where we fixed the value of $\mW$ as $\tau$.
% for iterative reweighting, the results looked horrible.
% DWT experiments for heavisine use weighted L1, but pcwreg and pcwpoly are not reweighted.

A brief summary of the results for our experiments is as follows. We observed that combining linear dynamic model with the $\ell_1$-norm regularization for sparse signal reconstruction provided much better signal reconstruction compared to the Kalman filter or $\ell_1$-regularized problems alone. The computational cost and execution time for $\ell_1$-homotopy is significantly smaller than that for SpaRSA. Average number of homotopy steps for updating the solution at every iteration ranges from 3 to 10, and average time for an update ranges from 5 to 13 milliseconds (the results in Fig.~\ref{fig:heavisine_results}--\ref{fig:pcwreg_results} are summed over $128$ iteration).

% \section{Conclusion}
% \TBC

%\vspace{-1ex}
\bibliographystyle{IEEEtran}
\bibliography{Streaming}

\end{document}